\documentclass[pra,aps,10pt,twocolumn,amsmath,amssymb]{revtex4-2}

\usepackage[english]{babel}
\usepackage{graphicx}% Include figure files
\usepackage{bm}% bold math
\usepackage{color}
\usepackage{slashed}%slashed D, k, A, etc.
\usepackage{latexsym}
\usepackage{appendix}
\usepackage{dsfont}

\newcommand{\dd}{d}
\newcommand{\ii}{i}
\newcommand{\ee}{e}

\newcommand{\me}{m_\mathrm{e}} %electron mass

\newcommand{\invf}{\raisebox{0.5em}{\rotatebox{180}{\scriptsize $\mathrm{F}$}}}

\newcommand{\supup}{\uparrow, \uparrow}
\newcommand{\supdown}{\uparrow, \downarrow}
\newcommand{\sdownup}{\downarrow, \uparrow}
\newcommand{\sdowndown}{\downarrow, \downarrow}

\begin{document}
 
\title{Dynamical Sauter-Schwinger pair creation process from Feynman perspective: Comparison of boundary- and initial-value approaches}
\author{J. Z. Kami\'nski} \email{Jerzy.Kaminski@fuw.edu.pl}
\author{A. Bechler} \email{Adam.Bechler@fuw.edu.pl}
\author{M. M. Majczak} \email{Mateusz.Majczak@fuw.edu.pl}
\author{K. Krajewska} \email{Katarzyna.Krajewska@fuw.edu.pl}

\affiliation{
Institute of Theoretical Physics, Faculty of Physics, University of Warsaw, Pasteura 5, 02-093 Warsaw, Poland}
\date{\today}

\begin{abstract}
We investigate the dynamical Sauter-Schwinger pair creation process from the vacuum by an electromagnetic background field using two alternative approaches.
The first one is based on the Feynman interpretation of positrons and the space-time description of Quantum Electrodynamics, which leads to the spin 
and momentum probability amplitudes expressed as the infinite Born series with respect to the background field. We demonstrate that in order to sum up 
this series exactly, the problem can be reduced to solving the Dirac equation with uniquely defined Feynman or anti-Feynman boundary conditions. 
The use of these boundary conditions leads to the results that are equivalent to the scattering matrix theory and consistent  with the worldline formalism. Alternative way of investigating 
the dynamical Sauter-Schwinger process consists in solving the Dirac equation with normalized initial (final) conditions. It is shown that this method follows 
from the suitably modified Feynman space-time approach, in which the Feynman propagators are replaced by the retarded (advanced) propagators. By doing 
so it is implicitly assumed that negative energy solutions describe electrons filling the Dirac sea (i.e., representing the Dirac vacuum) and 
the process of pair creation consists in the excitation of these electrons to the positive energy states. For both the boundary- and 
initial-value approaches the helicity-entangled momentum distributions are discussed and compared. Predictions of the two approaches are illustrated numerically 
for the homogeneous electric field pulse and for parameters such that for the spin summed up distributions both methods lead to nearly the same, although 
not identical, results. It is shown that even in such cases the spin- or helicity-resolved momentum distributions exhibit significant differences. 
\end{abstract}

\maketitle

\section{Introduction}
\label{sec:introduction}

We start our discussion by quoting the Bjorken and Drell textbook~\cite{bjorken1964relquantum}: 
\textit{
``The propagator approach to a relativistic quantum theory pioneered in 1949 by Feynman has provided a practical, as well as intuitively appealing, 
formulation of quantum electrodynamics and a fertile approach to a broad class of problems in the theory of elementary particles. The entire 
renormalization program, basic to the present confidence of theorists in the predictions of quantum electrodynamics, is in fact dependent on a Feynman 
graph analysis, as is also considerable progress in the proofs of analytic properties required to write dispersion relations. Indeed, one may go so 
far as to adopt the extreme view that the set of all Feynman graphs \underline{is} the theory''.
}
The subsequent development of relativistic quantum field theories, starting from Quantum Electrodynamics (QED) and culminating in the formulation 
of the Standard Model, seems to confirm this point of view (see, e.g.,~\cite{tHooft:1973wag,Veltman,itzykson1980quantumfield,Alvarez2012quantumfield,Kleiss,AuriaTrigiante}), 
at least from computational and practical perspectives. In fact, the first proof of the renormalizability of the  electroweak unification 
theory~\cite{Glashow,SalamWard,PhysRevLett.19.1264} utilized the Feynman diagrammatic approach (see, e.g.,~\cite{HooftVeltman} and references therein), 
as described for instance in~\cite{WeinbergBook}. Therefore, in this paper we also use the Feynman space-time 
approach~\cite{feymann1949positrons,feymann1949qed,feynman1998quantum} (for historical notes see, 
e.g.,~\cite{RevModPhys.58.449,SchweberQEDHistory,Wuthrich}). More specifically, we discuss what conclusions it leads to if applied to the dynamical Sauter-Schwinger 
process -- generalization of the pair creation process by a constant electric field, that was originally investigated by Sauter~\cite{sauter1931pairdirac} 
and Schwinger~\cite{schwinger1951gaugeinvariance}.

Dynamical Sauter-Schwinger pair creation is a fundamental quantum process in Strong Field QED (SFQED), an area investigating processes driven 
by electromagnetic background fields that are descibed by classical four-vector potentials \(A^{\mu}(x)\). These fields either modify existing QED processes (e.g., Compton or Mott scattering) 
or induce entirely new phenomena, such as a field-driven pair creation. 
Even though some SFQED processes have been studied just after the theoretical formulation of QED (see, 
e.g.,~\cite{sauter1931pairdirac,heisenberg1936diractheory,schwinger1951gaugeinvariance}), a revival of interest in this subject occurred 
with the discovery of very strong magnetic fields in the Universe (see, e.g.,~\cite{10.3389fphy.2014.00059,Enoto_2019} and references therein) and the development 
of experimental methods that allow generation of strong electromagnetic fields in laboratory conditions. The latter can be accomplished, for example, 
as a result of heavy ion collisions~\cite{Greiner1985QuantumElectrodynamics}, channeling of the high-energy charged particles 
(possibly protons or antiprotons from the Large Hadron Collider) through crystals or thin foils~\cite{RevModPhys.46.129}, or by amplifying and shortening the time duration of many laser 
pulses~\cite{PhysRevLett.104.220404}. Let us also mention a new direction related to the bound-free electron-positron pair production scenario discussed 
recently in~\cite{b9wk-hk6s}. The bibliography of this subject is nowadays enormous, hence, we cite only some selected books and 
review articles~\cite{fradkin1991vacuumquantum,grib1994vacuumquantum,Khalilov1996,Ritus,Nikishov,Rafelski1978FermionsAB,roshchupkin1996resonant,RevModPhys.78.309,salamin2006relativistic,ehlotzky2009fundqed,ruffini2010pairastro,dipiazza2012laserinter,GelisTanji2016,cajiao2019electron,sun2022pair,RevModPhys.94.045001,Brodin2022,fedetov2023qed,popruzhenko2023dynamics,sarri2025input}.

Historically (see, e.g.,~\cite{SchweberQEDHistory} and the Feynman's lecture in~\cite{FeynmanWeinberg1987}), the development of QED occurred in two steps, 
which were inextricably linked to the existence of negative energy solutions of the free Dirac equation and their physical interpretation. At first, 
the Dirac interpretation was adopted. In this approach, the negative energy solutions described electrons, which, thanks to the Pauli exclusion principle, 
filled the so-called Dirac sea~\cite{dirac1930theory}. Additionally, the absence of an electron in it was interpreted as a positron, following the 
discovery of this particle in cosmic rays by Anderson~\cite{PhysRev.43.491}. This point of view, which one can call the pre-Feynman approach, 
sufficiently well described many quantum processes (e.g., the Dirac~\cite{Dirac1930Annihilation} and Breit-Wheeler~\cite{breit1934collision} processes, 
or the creation of pairs by quantum tunneling, first investigated by Sauter~\cite{sauter1931pairdirac}) until radiative corrections began to be 
studied~\cite{SchweberQEDHistory}. Moreover, the Dirac interpretation was exclusively applicable to fermions. It was not applicable to bosons with zero 
spin, for which one also encounters negative energy solutions of the free Klein-Gordon equation. To solve these problems, Feynman interpreted negative 
energy solutions as those describing antiparticles, introduced a new type of Green's functions (also present in the earlier work by 
Tomonaga~\cite{10.1143/PTP.1.27}; see also~\cite{PhysRev.74.218} and references therein), which are now commonly called the Feynman 
propagators~\cite{bjorken1964relquantum}, and proposed the diagrammatic way of determining the radiative corrections. The Feynman approach led to 
the development of modern renormalizable QED, in which the Dirac sea no longer appears. Independently of Feynman research, using other equivalent methods,
Schwinger investigated the pair creation process in a constant electric field~\cite{schwinger1951gaugeinvariance}, arriving at results very close 
to those obtained by Sauter for weak electric fields (i.e., fields that are much smaller than the Schwinger value $\mathcal{E}_S$ defined below, although 
this quantity appeared already in the Sauter work~\cite{sauter1931pairdirac}). Let us emphasize, however, that the Schwinger analysis is more general. 
In particular, using this method and Feynman diagrams, one can incorporate radiative (see, e.g., 
Ref.~\cite{RitusEffectiveLagrangian,Ritus1986,WOS:A1985AEY1800001,DittrichGies}) and finite temperature (see, 
e.g.,~\cite{DittrichPLB,PhysRevD.19.2385,Bechler:1981an,KamPLB,WOS:A1981MH23000002,PhysRevD.61.085021,HATTORI2023104068}) corrections. 

The plan of this work is as follows. Based on Feynman original publications~\cite{feymann1949positrons,feymann1949qed,feynman1998quantum}, 
in Sec.~\ref{sec:spacetime} we apply the space-time approach to the problem of pair creation by an external electromagnetic field. This process 
is defined by Feynman diagrams, in which the free-fermion propagator is used that satisfies strictly defined asymptotic conditions in the far past 
and future. We show how the investigation of the pair creation process can be reduced to solving the Dirac equation with additional constraints 
imposed on possible solutions, that follow directly from the asymptotic conditions for the free-fermion propagator. Those constraints are known
as the Feynman and anti-Feynman boundary conditions, and have been introduced in Ref.~\cite{bialynicki1975quantumelectro}. They specify the asymptotic 
behavior of solutions of the Dirac equation corresponding to positive and negative energies. In Sec.~\ref{sec:Fworldline} we show that these boundary 
conditions are consistent with the so-called worldline formalism, which also follows directly from the space-time description of QED processes. 
Sec.~\ref{sec:spin} is devoted to the analysis of spin degrees of freedom of created pairs, where we also discuss how to define spin-entangled states 
and their corresponding probability amplitudes of pair creation. The case of homogeneous electric field pulses is considered in Sec.~\ref{sec:timeF}.

There is a number of methods which analyze the dynamical Sauter-Schwinger process by means of the Dirac equation (see, for 
instance, Refs.~\cite{avetissian2002pair,blinne2014pair,blinne2016wigner} or the so-called Computational Quantum Field 
Theory method~\cite{doi:10.1080/00107510903450559,li2023phasepair,xylm-fnt7,Li_2026}), using the initial conditions for the solutions of the Dirac 
equation. Also, the broadly applied formalisms based on the Quantum Kinetic Equations (QKE) (see, e.g.,~\cite{PhysRevResearch.6.043009}) or the 
Dirac-Heisenberg-Wigner function (DHW) \cite{bialynicki1991diracvacuum} follow from the initial-value Dirac method, which for homogeneous 
time-dependent electric pulses was demonstrated in Ref.~\cite{Bechler_Krajewska_CajiaoVelez_Kaminski_2023}. In this regard, we show in Sec.~\ref{sec:ret} that 
the method based on solving the Dirac equation with the initial condition is equivalent to the modified Feynman space-time theory provided that
the Feynman propagator is replaced by the retarded propagator. Although mathematically correct, this is not consistent with the interpretation of 
the negative energy solutions as antifermions, as already noted by Feynman in~\cite{feynman1998quantum}. However, if the contributions of the negative 
energy parts of the Feynman and retarded propagators are small (and this usually happens for electromagnetic field strengths much smaller then 
the Schwinger values, either of the electric or magnetic fields, or for sufficiently high energies of created pairs) then this two approaches 
lead to nearly identical (although, not exactly the same) spin summed up momentum distributions~\cite{PhysRevD.110.116025}. In Sec.~\ref{sec:timeR} 
the case of homogeneous electric field pulses is discussed, whereas the spin-resolved amplitudes are considered in Sec.~\ref{sec:spininitialvalue}. 
In order for our presentation to be more complete, we also summarize in Sec.~\ref{sec:qke} results related to the QKE approach. 
At this point, it is worth noting that in two works by Ritus~\cite{Ritus} and Nikishov~\cite{Nikishov}, considered to be fundamental for 
the development of SFQED, the concept of QKE does not appear.

Sec.~\ref{sec:num} is devoted to numerical analysis in which predictions of the boundary- and initial-value approaches are compared for homogeneous 
electric field pulses. The motivation for considering such a simplified case is fourfold. First, analyzing electric field pulses that depend only on time 
reduces the problem to numerical solution of a system of ordinary differential equations. This, in turn, allows the numerical calculations to be 
performed with essentially an arbitrary precision. This plays a significant role when comparing the two methods, as it eliminates random numerical 
errors that can affect the interpretation of the results. Moreover, the results obtained here can serve as a reference and verification point for 
different approximations used in other approaches, for instance those applied in the worldline formalism. Second, analogues of the dynamical 
Sauter-Schwinger process are also analyzed in other areas of physics, such as condensed matter physics (see, 
e.g.,~\cite{RevModPhys.81.109,PhysRevD.108.116007,PhysRevLett.124.110403,PhysRevD.99.016025,suster2020dynamical}) or photonics (see, 
e.g.,~\cite{PhysRevLett.109.110401}). Therefore, the investigations presented here can be adapted to problems arising there. Third, experimental 
verification of the Sauter-Schwinger process from the vacuum is still rather a dream for researchers working in this area, so studies focus 
primarily on fundamental aspects rather than comparing theoretical predictions with experimental results. Finally, the development of laser technology 
suggests that by focusing multiple light beams propagating in different directions, it is possible to amplify the electric component and simultaneously 
weaken (or even suppress) the magnetic component of laser pulses in certain regions. Therefore, one can expect that exactly in these regions 
the pair creation via the Sauter-Schwinger process should be dominant. For this reason, in our numerical analysis we consider an electric field pulse 
that is a superposition of three circularly polarized pulses with polarization planes perpendicular to each other and, in general, of different helicities and carrier envelope 
phases. By doing this we also eliminate an accidental disappearance of some spin-resolved probability amplitudes, which makes the spin analysis of 
the process more complete. In Sec.~\ref{sec:QKEversusInitial}, we compare the spin-momentum distributions of created electrons using methods based 
on QKE and solutions of the Dirac equation with the initial condition. We show that these distributions are identical if summed over the spin degrees 
of freedom of the initial negative energy states. This means that the analysis of the Dirac equation with the initial condition provides a more 
comprehensive understanding of the process of electron creation from the Dirac sea. Sec.~\ref{sec:BoundaryversusInitial} is devoted to the comparison 
of approaches based on solutions of the Dirac equation with boundary and initial conditions. Our analysis focuses on the situation in which both 
methods predict nearly the same (although never identical) momentum distributions of created electrons summed over the spin degrees of freedom. 
However, it turns out that even in this case, the helicity-resolved momentum distributions differ significantly for these two approaches. The same 
effect is also observed in the helicity-entangled momentum distributions analyzed in Sec.~\ref{sec:Entanglment}.

In numerical analysis, we use the relativistic units in which $\hbar=c=\me=|e|=1$ and $e=-|e|$, where $\me$ and $e$ are the electron rest  mass and charge. The Schwinger value for the electric field strength, $\mathcal{E}_S$, is defined such that $\me c^2=|e|\mathcal{E}_S\lambdabar_C$, where $\lambdabar_C=\hbar/\me c$ is the reduced Compton wavelength. Additionally, the Compton time is defined as $t_C=\hbar/\me c^2$. For the $\gamma$ matrices we apply the Dirac representation, use the Feynman notation $\slashed{a}=a_\mu\gamma^\mu$, and the metric $(+,-,-,-)$. In analytical formulas we put $\hbar=1$ but keep explicitly $\me$, $e$, and $c$.

\begin{figure}
\includegraphics[width=7.5cm]{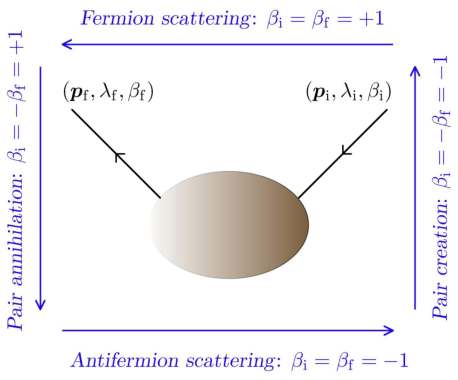}
\caption{Schematic drawing of the Feynman diagram for the quantum process with two external lines representing the incoming fermion of the momentum 
$\bm{p}_\mathrm{i}$ and the spin polarization $\lambda_\mathrm{i}=\pm$, and the outgoing fermion of the momentum $\bm{p}_\mathrm{f}$ and the spin 
polarization $\lambda_\mathrm{f}=\pm$. Both particles are on the mass shell, i.e., $p^0_\mathrm{i}=\sqrt{\bm{p}^2_\mathrm{i}+m^2c^2}$ and 
$p^0_\mathrm{f}=\sqrt{\bm{p}^2_\mathrm{f}+m^2c^2}$, where $m$ is the fermion rest mass. The third label, $\beta_\mathrm{i}=\pm$ or 
$\beta_\mathrm{f}=\pm$, indicates whether we deal with particles ($+$) or antiparticles ($-$). The fermion line arrows discriminate particles 
and antiparticles, i.e., if the arrow is aligned with the time-arrow (which, for the given process, is represented by the blue arrow), the line represents a particle, 
otherwise it represents an antiparticle. Since particles and antiparticles have opposite electric charges and we assume that the total electric 
charge is conserved during the time-evolution, therefore, for all possible time-arrows the $\beta$ labels have to be properly selected, as shown
in the figure. The central ellipse indicates the space-time region in which the classical background field does not vanish and in which all possible 
virtual fermion pairs are created and annihilated, as well as the virtual photons are emitted and absorbed by fermion lines. In order not to complicate 
the picture, possible processes with the absorption and emission of real photons are neglected. This diagram is a modification of Fig.~3 from 
the original Feynman paper~\cite{feymann1949positrons}.
}
\label{diaeggmmf}
\end{figure}

\section{Space-time description of the electron-positron pair creation}
\label{sec:spacetime}

Consider a quantum process described by a Feynman diagram, presented in Fig.~\ref{diaeggmmf}, with one incoming and one outgoing fermion line. 
The free state of the incoming fermion line is labeled by three indexes, $(\bm{p}_{\mathrm{i}},\lambda_{\mathrm{i}},\beta_{\mathrm{i}})$, of 
which $\bm{p}_{\mathrm{i}}$ denotes the fermion momentum, $\lambda_{\mathrm{i}}=\pm$ stands for the spin polarization, and $\beta_{\mathrm{i}}=\pm$ 
distinguishes particles ($+$) from antiparticles ($-$). Similarly, for the outgoing line, we have 
$(\bm{p}_{\mathrm{f}},\lambda_{\mathrm{f}},\beta_{\mathrm{f}})$. The probability amplitude of this process is defined by the formula
\begin{align} \label{f1}
\mathcal{A}(\bm{p}_{\mathrm{i}},\lambda_{\mathrm{i}},\beta_{\mathrm{i}}&\rightarrow \bm{p}_{\mathrm{f}},\lambda_{\mathrm{f}},\beta_{\mathrm{f}}) \\
&= \int \dd^4x\dd^4y\bar{\chi}^{(\beta_{\mathrm{f}})}_{\bm{p}_{\mathrm{f}},\lambda_{\mathrm{f}}}(y)M(y,x)\chi^{(\beta_{\mathrm{i}})}_{\bm{p}_{\mathrm{i}},\lambda_{\mathrm{i}}}(x), \nonumber
\end{align}
where the matrix $4\times 4$, $M(x,y)$, describes both virtual processes and the interaction with the external classical electromagnetic field in the shaded region. The free-fermion states have in general the following form,
\begin{equation}\label{f2}
\chi^{(\beta)}_{\bm{p},\lambda}(x)=u^{(\beta)}_{\bm{p},\lambda}\ee^{-\ii\beta p\cdot x},
\end{equation}
in which $p^0=\sqrt{\bm{p}^2+m^2c^2}>0$ and
\begin{equation}\label{f3}
\!\!  u^{(+)}_{\bm{p},\lambda}\! =\! N\! 
\begin{pmatrix}\! \! 
(p^0\! +\! mc)\chi^{(+)}_{\lambda}
\cr
\bm{\sigma}\cdot\bm{p}\chi^{(+)}_{\lambda}
\!\!  \end{pmatrix}
\!,
u^{(-)}_{\bm{p},\lambda}\! =\! N\! 
\begin{pmatrix}\! \!    
\bm{\sigma}\cdot\bm{p}\chi^{(-)}_{\lambda}
\cr
(p^0\! +\! mc)\chi^{(-)}_{\lambda}
\!\!  \end{pmatrix}
\! ,
\end{equation}
where $\bm{\sigma}$ are the Pauli matrices, whereas the normalization factor equals
\begin{equation}\label{f4}
N=\frac{1}{\sqrt{2p^0(p^0+mc)}}.
\end{equation}
Moreover, the spinors $\chi^{(\beta)}_{\lambda}$ have the form,
\begin{equation}\label{f3a}
\chi^{(\beta)}_{+}\! =\!
\begin{pmatrix}
\cos\frac{\theta_{\beta}}{2}\ee^{-\ii\varphi_{\beta}/2}
\cr
\sin\frac{\theta_{\beta}}{2}\ee^{\ii\varphi_{\beta}/2}
\end{pmatrix}
\!,\!
\chi^{(\beta)}_{-}\! =\! 
\begin{pmatrix}
-\sin\frac{\theta_{\beta}}{2}\ee^{-\ii\varphi_{\beta}/2}
\cr
\cos\frac{\theta_{\beta}}{2}\ee^{\ii\varphi_{\beta}/2}
\end{pmatrix}
,
\end{equation}
where $\theta_{\beta}$ and $\varphi_{\beta}$ are the polar and azimuthal angles that define the spin quantization axes, in general different for 
the incoming and outgoing lines. Note that intentionally the label $\lambda$ is called the spin polarization and not the spin projection on the selected 
axis, as for antiparticles the projections of the spin are equal to $-\lambda\hbar/2=\beta\lambda\hbar/2$. We use states defined by Eqs.~\eqref{f2} 
and~\eqref{f3} mostly as the computational basis.

\begin{figure}
\includegraphics[width=7.5cm]{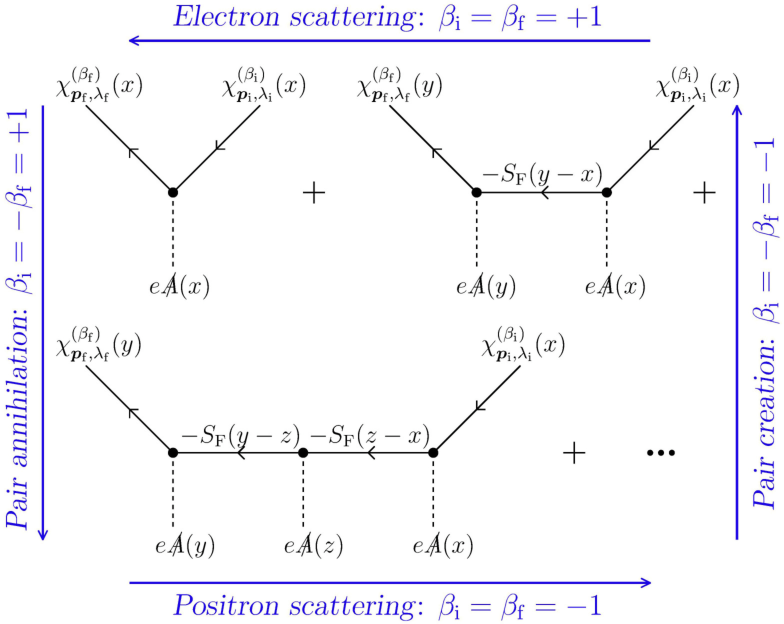}
\caption{Feynman diagrams representing four possible QED processes with electrons and positrons in the tree approximation and without emission 
or absorption of real photons. For the pair creation and radiationless pair annihilation processes also the mutual electromagnetic interaction 
between created (annihilated) electrons and positrons in the far future (past) is neglected. Note that the minus sign in front of $S_{\mathrm{F}}$ 
follows from our definition~\eqref{fff3}.
}
\label{diagrammm}
\end{figure}

Within the scattering matrix formalism, when neglecting radiative corrections but exactly accounting for interactions with an external electromagnetic field, 
the matrix $M(y,x)$ from Eq.~\eqref{f1} takes the form (see, e.g.,~\cite{PhysRevD.110.116025}),
\begin{equation}\label{fff1}
M(y,x)\! =\! -\ii[e\slashed{A}(x)\delta^{(4}(x\! -\! y)
-e\slashed{A}(y)K_{\mathrm{F}}(y,x)e\slashed{A}(x)].
\end{equation}
Here, the exact Feynman propagator, $K_{\mathrm{F}}(y,x)$, satisfies the following integral equations, 
\begin{align}\label{fff2}
\!\!\!\!  K_{\mathrm{F}}(y,x)\! =&S_{\mathrm{F}}(y-x)\! -\!\! \int\! \dd^4z S_{\mathrm{F}}(y-z)e\slashed{A}(z)K_{\mathrm{F}}(z,x), \\
=&S_{\mathrm{F}}(y-x)\! -\!\! \int\! \dd^4z K_{\mathrm{F}}(y,z)e\slashed{A}(z)S_{\mathrm{F}}(z-x), \nonumber
\end{align}
which can also be expressed as an infinite Born series.
The free-fermion Feynman propagator, $S_{\mathrm{F}}(z)$, is defined by the integral (with an infinitesimally small and positive $\varepsilon$, 
which in~\cite{PhysRevD.2.1191} is called 'the correct $\ii\varepsilon$ prescription' [see the text below equation (11)]),
\begin{equation}\label{fff3}
S_{\mathrm{F}}(z)=-\int\frac{\dd^4q}{(2\pi)^4}\frac{\slashed{q}+mc}{q^2-(mc)^2+\ii\varepsilon}\ee^{-\ii q\cdot z},  
\end{equation}
and which is a particular solution of the inhomogeneous Dirac equation,
\begin{equation}\label{fff4}
\Bigl(-\ii\gamma^{\mu}\frac{\partial}{\partial z^{\mu}}+mc\Bigr)S_{\mathrm{F}}(z)=\delta^{(4)}(z).  
\end{equation}
This means that the full Feynman propagator~\eqref{fff2} is also a solution of the inhomogeneous Dirac equation,
\begin{equation}\label{fff4a}
\Bigl(-\ii\gamma^{\mu}\frac{\partial}{\partial y^{\mu}}+e\slashed{A}(y)+mc\Bigr)K_{\mathrm{F}}(y,x)=\delta^{(4)}(y-x).  
\end{equation}
Moreover, by expanding $K_{\mathrm{F}}(y,x)$ in the infinite Born series, one can represent the probability amplitude~\eqref{f1}, 
up to the global phase factor $-\ii$, by the Feynman diagrams presented in Fig.~\ref{diagrammm} for electrons and positrons. 
At this point let us remark that $S_{\mathrm{F}}$ and $K_{\mathrm{F}}$ are not the only exact solutions of Eqs.~\eqref{fff4} and~\eqref{fff4a}. 
This fact is crucial for our further discussion, where we compare the consequences of the Feynman theory with other approaches.
Note also that this formulation assumes the equivalence of the S-matrix and Feynman space-time approaches, a relationship discussed by Dyson~\cite{dyson1949radiation,dyson1949s} and corroborated 
in literature by Nikishov~\cite{Nikishov1986}.

Furthermore, with our choice of the factor \eqref{f4}, the free-fermion states are normalized such that
\begin{equation}\label{f5}
\!\!\! \int\!\dd^3x\, \bar{\chi}^{(\beta)}_{\bm{p},\lambda}(x) \gamma^0\! \chi^{(\beta')}_{\bm{p}',\lambda'}(x)\! =\!
(2\pi)^3\delta^{(3)}\!(\bm{p}-\bm{p}')\delta_{\lambda,\lambda'}\delta_{\beta,\beta'}.
\end{equation}
Hence, the free-fermion Feynman propagator becomes
\begin{equation}\label{fh1}
S_{\mathrm{F}}(y-x)\! =\ii\! \sum_{\lambda,\beta=\pm}\!\! \beta\theta(\beta(y^0-x^0))\!\! \int\!\! \frac{\dd^3q}{(2\pi)^3}
\chi^{(\beta)}_{\bm{q},\lambda}(y)\bar{\chi}^{(\beta)}_{\bm{q},\lambda}(x),
\end{equation}
where $\theta(t)$ is the Heaviside function equal to 1 for positive $t$ and 0 otherwise. In consequence, one obtains the equations,
\begin{align}\label{f6a}
\!\!\!\!  \int\! \dd^3x S_{\mathrm{F}}(y-x)\gamma^0 \chi^{(\beta)}_{\bm{p},\lambda}(x)\! =\! \ii\beta\theta(\beta(y^0-x^0))
\chi^{(\beta)}_{\bm{p},\lambda}(y) 
\end{align}
and
\begin{align}\label{f6b}
\!\!\!\!   \int\! \dd^3y \bar{\chi}^{(\beta)}_{\bm{p},\lambda}(y)\gamma^0 S_{\mathrm{F}}(y-x)\! =\! \ii\beta\theta(\beta(y^0-x^0))\bar{\chi}^{(\beta)}_{\bm{p},\lambda}(x)
, 
\end{align}
which will be used below.

In our further discussion, we consider electron-positron ($e^-e^+$) pair creation (hence, $m=\me$ and $e$ is the electron charge) 
from the vacuum by a strong space and time dependent external electromagnetic field, that is described by the vector potential $A_\mu(x)$, 
with $x=(x^0,{\bm x})=(ct,\bm{x})$. We assume that both the electromagnetic field and the vector potential vanish for $t\rightarrow -\infty$ and 
$t\rightarrow\infty$. While the charged particles move in the external field, their mutual electromagnetic interactions 
are disregarded. Moreover, the symbol
\begin{equation}\label{f7}
\mathcal{A}(\bm{p}_-,\lambda_-;\bm{p}_+,\lambda_+)=\mathcal{A}(\bm{p}_{+},\lambda_{+},-\rightarrow \bm{p}_{-},\lambda_{-},+)
\end{equation}
will denote the probability amplitude of the electron-positron pair creation. Hence, $\bm{p}_-$ and $\lambda_-$ are the electron momentum and spin polarization, and $\bm{p}_+$ and $\lambda_+$ apply to positrons.

Further analysis of the probability amplitude~\eqref{f7} can follow via two paths. First, one can use the exact form of the Feynman propagator 
$K_{\mathrm{F}}(y,x)$, often represented in terms of a functional integral. 
Evaluating the amplitude
requires approximations
-- mostly (but not only) the saddle point approximation, which forms the basis of the worldline formalism~\cite{schubert2001perturbative,PhysRevD.72.105004,PhysRevD.73.065028,PhysRevD.84.125023,PhysRevD.104.105014,PhysRevD.107.056019}. 
While it is known that the worldline approach connects directly to the scattering matrix (see, e.g.,~\cite{PhysRevD.107.056019}), for completeness we will discuss it in 
Sec.~\ref{sec:Fworldline}. The second path, which is the focus of our analysis, reduces the problem to solving a differential equation. 
Using this framework, one can analyze pair 
creation driven by time-dependent electric field pulses to arbitrary precision, primarily through numerical methods. 
Regardless of the approach, they both are based on the Feynman space-time theory.

To proceed, let us recall (see, e.g.,~\cite{PhysRevD.110.116025}) that the form of $M(y,x)$ defined by Eq.~\eqref{fff1} allows us to introduce 
the so-called Feynman state,
\begin{equation}\label{f8}
\psi^{(-)}_{\mathrm{F};\bm{p}_+,\lambda_+}(y)=\chi^{(-)}_{\bm{p}_+,\lambda_+}(y)-\! \int\! \dd^4x K_{\mathrm{F}}(y,x)e\slashed{A}(x)\chi^{(-)}_{\bm{p}_+,\lambda_+}(x),
\end{equation}
such that
\begin{equation}\label{f9}
\mathcal{A}(\bm{p}_-,\lambda_-;\bm{p}_+,\lambda_+)=\! -\ii\! \int\!  \dd^4y \bar{\chi}^{(+)}_{\bm{p}_{-},\lambda_{-}}(y)e\slashed{A}(y)\psi^{(-)}_{\mathrm{F};\bm{p}_+,\lambda_+}(y).
\end{equation}
It is straightforward to show that the Feynman state \eqref{f8} satisfies the Dirac equation,
\begin{equation}\label{f10}
\Bigl(-\ii\gamma^{\mu}\frac{\partial}{\partial y^{\mu}}+e\slashed{A}(y)+\me c\Bigr)\psi^{(-)}_{\mathrm{F};\bm{p}_+,\lambda_+}(y)=0.  
\end{equation}
However, in order to uniquely determine its solution we have to impose proper initial (final) or boundary conditions. The latter conditions are not arbitrary 
and have to follow from the definition~\eqref{f8} and from the asymptotic behavior of the Feynman propagator $K_{\mathrm{F}}(y,x)$. 
Namely, by applying Eqs.~\eqref{fff2} and~\eqref{fh1}, one obtains the asymptotic form of the Feynman propagator at $y^0\rightarrow \pm\infty$ 
and fixed $x^0$,
\begin{equation}\label{f11}
K_{\mathrm{F}}(y,x)=\ii\!\! \sum_{\lambda,\beta=\pm}\!\! \beta\theta(\beta(y^0-x^0))\! \int\! \frac{\dd^3q}{(2\pi)^3}
\chi^{(\beta)}_{\bm{q},\lambda}(y)\bar{\psi}^{(\beta)}_{\mathrm{\invf};\bm{q},\lambda}(x).
\end{equation}
Here,
\begin{equation}\label{f12}
\psi^{(\beta)}_{\mathrm{\invf};\bm{q},\lambda}(x)=\chi^{(\beta)}_{\bm{q},\lambda}(x)
-\int\dd^4z K_{\mathrm{\invf}}(x,z)e\slashed{A}(z)\chi^{(\beta)}_{\bm{q},\lambda}(z),
\end{equation}
or
\begin{equation}\label{f12a}
\bar{\psi}^{(\beta)}_{\mathrm{\invf};\bm{q},\lambda}(x)=\bar{\chi}^{(\beta)}_{\bm{q},\lambda}(x)
-\int\dd^4z \bar{\chi}^{(\beta)}_{\bm{q},\lambda}(z)e\slashed{A}(z)K_{\mathrm{F}}(z,x),
\end{equation}
is for $\beta=+$ the so-called anti-Feynman state~\cite{bialynicki1975quantumelectro,PhysRevD.110.116025}, whereas
\begin{equation}\label{f13}
K_{\mathrm{\invf}}(x,z)=\gamma^0[K_{\mathrm{F}}(z,x)]^{\dagger}\gamma^0
\end{equation}
is the anti-Feynman propagator. Note, that the explicit form of $\bar{\psi}^{(\beta)}_{\mathrm{\invf};\bm{q},\lambda}(x)$ results directly 
from Eqs.~\eqref{fff2} and~\eqref{fh1}. It follows from these equations that $\psi^{(-)}_{\mathrm{F};\bm{p}_+,\lambda_+}(y)$, being solution of the Dirac 
equation, also has to satisfy the following boundary conditions: for $y^0\rightarrow\infty$,
\begin{align}\label{fb1a}
\psi^{(-)}_{\mathrm{F};\bm{p}_+,\lambda_+}(y)=&
\chi^{(-)}_{\bm{p}_+,\lambda_+}(y) \\
+&\sum_{\lambda=\pm}\int\frac{\dd^3q}{(2\pi)^3}C^{(+)}_{\mathrm{F};\bm{p}_+,\lambda_+}(\bm{q},\lambda)\chi^{(+)}_{\bm{q},\lambda}(y), \nonumber
\end{align}
and for $y^0\rightarrow -\infty$,
\begin{align}\label{fb1b}
\psi^{(-)}_{\mathrm{F};\bm{p}_+,\lambda_+}(y)=&
\chi^{(-)}_{\bm{p}_+,\lambda_+}(y) \\
+&\sum_{\lambda=\pm}\int\frac{\dd^3q}{(2\pi)^3}C^{(-)}_{\mathrm{F};\bm{p}_+,\lambda_+}(\bm{q},\lambda)\chi^{(-)}_{\bm{q},\lambda}(y), \nonumber 
\end{align}
where
\begin{equation}\label{f14}
C^{(\beta)}_{\mathrm{F};\bm{p}_+,\lambda_+}(\bm{q},\lambda)=-\ii\beta\int\dd^4x \bar{\psi}^{(\beta)}_{\mathrm{\invf};\bm{q},\lambda}(x)e\slashed{A}(x)\chi^{(-)}_{\bm{p}_+,\lambda_+}(x).
\end{equation}
Thus, the boundary conditions under which the Dirac equation has to be solved are as follows: (i) in the remote past there is a superposition of only negative energy solutions of the Dirac 
equation; (ii) in the far future there is a superposition of positive energy solutions and the well-defined negative energy solution, that 
corresponds to the created positron. These are exactly the Feynman boundary conditions introduced in Ref.~\cite{bialynicki1975quantumelectro}.
They were also discussed later in Ref.~\cite{Nikishov}, and applied to the dynamical Sauter-Schwinger process~\cite{PhysRevD.110.116025}, where it 
was demonstrated that
\begin{equation}\label{f15}
\mathcal{A}(\bm{p}_-,\lambda_-;\bm{p}_+,\lambda_+)=C^{(+)}_{\mathrm{F};\bm{p}_+,\lambda_+}(\bm{p}_-,\lambda_-).
\end{equation}
This equation, according to the presented here analysis, immediately follows from~\eqref{f12a}. It also resembles the similar one known from scattering 
theory~\cite{goldberger1964collision,newton1982scattering} in which the scattering amplitude can be expressed either by the solution with the outgoing 
spherical waves (for the considered problem, by the solution satisfying the Feynman boundary condition) or by the complex conjugate solution with 
the incoming spherical waves (in our case, the solution fulfilling the anti-Feynman boundary conditions). 

The above analysis shows that by fixing the momentum and spin polarization of the created positron state in the future we uniquely determine both 
the generated electron wave packet in the future, with the profile $C^{(+)}_{\mathrm{F};\bm{p}_+,\lambda_+}(\bm{q},\lambda)$, and the wave packet 
of the negative energy states in the past (we intentionally do not call them positrons as there are no real particles in the past), with the profile 
$C^{(-)}_{\mathrm{F};\bm{p}_+,\lambda_+}(\bm{q},\lambda)$. This is why one can say that in the pair creation from the vacuum by external electromagnetic 
field it is the future that determines the past, and not \textit{vice versa}. Note that this statement cannot be confused with the final-value approach. 
In fact, the latter is equivalent to the initial-value approach, as it consists in propagating the well-defined final state backward in time (see, 
e.g.,~\cite{PhysRevD.94.065024}). This is in contrast to the boundary-value approach in which both the initial and final states are \textit{a priori} 
not known, but only parts of them are precisely defined. To be more specific, for the Feynman solution only the negative energy 
part is well-defined in the future, whereas in the past the positive energy part has to be equal to 0.

Since in QED there is the symmetry between electrons and positrons, and it is only our choice to call electrons as particles, a similar 
prescription should exists if one fixes the electron momentum and spin polarization. Indeed, such case was discussed in Ref.~\cite{PhysRevD.110.116025} 
and it follows from the formula,
\begin{equation}\label{f16}
[\mathcal{A}(\bm{p}_-,\lambda_-;\bm{p}_+,\lambda_+)]^*\!=\ii\! \int \!\dd^4y \bar{\chi}^{(-)}_{\bm{p}_{+},\lambda_{+}}\! (y)e\slashed{A}(y)\psi^{(+)}_{\mathrm{\invf};\bm{p}_-,\lambda_-}\! (y).
\end{equation}
Now we deal with the anti-Feynman state for which electrons in the future have the well-defined momentum and spin polarization. 
Therefore, in order to distinguish between these two cases we introduce two different notations for, in fact, the same amplitude,
\begin{align}\label{f17}
\mathcal{A}(\bm{p}_-,\lambda_-;\bm{p}_+,\lambda_+)&= \mathcal{A}^{(+)}_{\bm{p}_+,\lambda_+}(\bm{p}_-,\lambda_-) \\
&=\mathcal{A}^{(-)}_{\bm{p}_-,\lambda_-}(\bm{p}_+,\lambda_+) \nonumber .
\end{align}
Here, $\mathcal{A}^{(+)}_{\bm{p}_+,\lambda_+}(\bm{p}_-,\lambda_-)$ means that the positron momentum and spin polarization are fixed, whereas
$\mathcal{A}^{(-)}_{\bm{p}_-,\lambda_-}(\bm{p}_+,\lambda_+)$ is for the fixed state of created electron.

\bigskip

\subsection{Connections with worldline formalism}
\label{sec:Fworldline}

As it was discussed by Feynman (see, e.g., Ref.~\cite{feynman1998quantum}), Eq.~\eqref{f1} can be put in the more compact form,
\begin{widetext}
\begin{equation} \label{ff1}
\mathcal{A}(\bm{p}_{-},\lambda_{-}; \bm{p}_{+},\lambda_{+}) 
=\ii \lim_{y^0\rightarrow +\infty}\lim_{x^0\rightarrow +\infty}\int \dd^3x\dd^3y\bar{\chi}^{(+)}_{\bm{p}_{-},\lambda_{-}}(y)\gamma^0K_{\mathrm{F}}(y,x)\gamma^0\chi^{(-)}_{\bm{p}_{+},\lambda_{+}}(x),
\end{equation}
or, equivalently,
\begin{equation} \label{ff1a}
[\mathcal{A}(\bm{p}_{-},\lambda_{-}; \bm{p}_{+},\lambda_{+})]^* 
=-\ii \lim_{y^0\rightarrow +\infty}\lim_{x^0\rightarrow +\infty}\int \dd^3x\dd^3y\bar{\chi}^{(-)}_{\bm{p}_{+},\lambda_{+}}(x)\gamma^0K_{\mathrm{\invf}}(x,y)\gamma^0\chi^{(+)}_{\bm{p}_{-},\lambda_{-}}(y).
\end{equation}
Indeed, inserting 
\begin{equation}\label{fg1}
K_{\mathrm{F}}(y,x)=S_{\mathrm{F}}(y-x)-\int\dd^4zS_{\mathrm{F}}(y-z)e\slashed{A}(z)S_{\mathrm{F}}(z-x)
+\int\dd^4z\dd^4wS_{\mathrm{F}}(y-z)e\slashed{A}(z)K_{\mathrm{F}}(z,w)e\slashed{A}(w)S_{\mathrm{F}}(w-x)
\end{equation}
\end{widetext}
into Eq.~\eqref{ff1}, and applying~\eqref{f6a} and~\eqref{f6b} with the corresponding time limits, we derive the amplitude that follows directly 
from the scattering matrix formalism (see, e.g.,~\cite{feynman1998quantum,PhysRevD.107.056019}). In fact, we could start our discussion from the 
formula~\eqref{ff1} and arrive at Eq.~\eqref{f1}. Note also that, according to the Feynman formulas~\eqref{ff1} and~\eqref{ff1a}, 
the total electric charge is conserved, and equal to 0.

From the time-limits in Eq.~\eqref{ff1} we learn that both electron and positron states are defined in the future, as it should be for the pair 
creation from the vacuum. This follows directly from the Feynman boundary conditions imposed on the propagators. Thus, the worldline formalism 
(for which the Feynman theory is the starting point, if applied to QED processes) agrees with the scattering matrix theory as well as with the approach 
in which the Dirac equation is solved with the Feynman (or anti-Feynman) boundary conditions.

\subsection{Spin-resolved amplitudes}
\label{sec:spin}

The aim of this section is to analyze the spin degrees of freedom for pair creation. For this purpose, in order not to complicate the notation, 
let us introduce the short-hand symbol,
\begin{equation}\label{s1}
\mathcal{A}_{\lambda_-,\lambda_+}=\mathcal{A}(\bm{p}_-,\lambda_-;\bm{p}_+,\lambda_+),
\end{equation}
in which the momentum degrees of freedom are neglected. As we have already noted above, the symbols $\lambda_-$ and $\lambda_+$ discriminate 
between different spin polarizations. They are not always equal to the spin projections onto some particular axes, as it will be discussed below. 

From the purely practical point of view, we will carry out calculations for the particular spin polarizations defined by the angles 
$\theta_\beta=\varphi_\beta=0$ in Eq.~\eqref{f3}. Let us denote these probability amplitudes by $\mathcal{A}^{(0)}_{\lambda_-,\lambda_+}$. 
From the general formula~\eqref{f1} we learn that, in order to determine the probability amplitudes for arbitrary spin polarizations, 
it is sufficient to perform the matrix multiplication,
\begin{equation}\label{s2}
\begin{pmatrix}
\mathcal{A}_{+,+} \cr \mathcal{A}_{+,-} \cr \mathcal{A}_{-,+} \cr \mathcal{A}_{-,-} 
\end{pmatrix}
=\Biggl[\!\!
\begin{pmatrix}
c_+^* & s_+^* \cr
-s_+  & c_+
\end{pmatrix}
\otimes
\begin{pmatrix}
c_-     & s_- \cr
-s_-^*  & c_-^*
\end{pmatrix}
\!\!\Biggr]
\begin{pmatrix}
\mathcal{A}^{(0)}_{+,+} \cr \mathcal{A}^{(0)}_{+,-} \cr \mathcal{A}^{(0)}_{-,+} \cr \mathcal{A}^{(0)}_{-,-} 
\end{pmatrix}.
\end{equation}
Here, the symbol $\otimes$ stands for the Kronecker product of two matrices~\cite{Solve,Nielsen} and 
$c_{\beta}=\cos(\theta_{\beta}/2)\ee^{-\ii\varphi_{\beta}/2}$, $s_{\beta}=\sin(\theta_{\beta}/2)\ee^{\ii\varphi_{\beta}/2}$, separately for electrons
($\beta=+$) and positrons ($\beta=-$). Since $|c_{\beta}|^2+|s_{\beta}|^2=1$, both matrices in Eq.~\eqref{s2} are unitary and the spin summed up probability distribution is independent of the choice of the spin quantization axis. 

Let us now consider the spin projections which are denoted by the arrows up and down. The corresponding spinors \eqref{f3} are equal to,
\begin{equation}\label{s3}
\chi^{(+)}_{\uparrow}=\chi^{(+)}_+,
\chi^{(+)}_{\downarrow}=\chi^{(+)}_-,
\chi^{(-)}_{\uparrow}=-\chi^{(-)}_-,
\chi^{(-)}_{\downarrow}=\chi^{(-)}_+,
\end{equation}
which leads to the probability amplitudes for the spin projections,
\begin{align}\label{s4}
&\mathcal{A}_{\supup}=-\mathcal{A}_{+,-},\quad
\mathcal{A}_{\supdown}=\mathcal{A}_{+,+},
\\
&\mathcal{A}_{\sdownup}=-\mathcal{A}_{-,-},\quad
\mathcal{A}_{\sdowndown}=\mathcal{A}_{-,+}. \nonumber
\end{align}
If the spin quantization axes are the same (i.e., defined by the same polar and azimuthal angles, $\theta=\theta_{\beta}$ and $\varphi=\varphi_{\beta}$, respectively) then 
\begin{equation}\label{s2aa}
\begin{pmatrix}
\mathcal{A}_{+,+} \cr \mathcal{A}_{+,-} \cr \mathcal{A}_{-,+} \cr \mathcal{A}_{-,-} 
\end{pmatrix}
=\mathbb{S}(\theta,\varphi)
\begin{pmatrix}
\mathcal{A}^{(0)}_{+,+} \cr \mathcal{A}^{(0)}_{+,-} \cr \mathcal{A}^{(0)}_{-,+} \cr \mathcal{A}^{(0)}_{-,-} 
\end{pmatrix},
\end{equation}
where
\begin{widetext}
\begin{equation}\label{ss2}
\mathbb{S}(\theta,\varphi)
=\frac{1}{2}
\begin{pmatrix}
1+\cos\theta & \sin\theta\ee^{\ii\varphi} & \sin\theta\ee^{-\ii\varphi} & 1-\cos\theta \cr
-\sin\theta & (1+\cos\theta)\ee^{\ii\varphi} & -(1-\cos\theta)\ee^{-\ii\varphi} & \sin\theta \cr
-\sin\theta & -(1-\cos\theta)\ee^{\ii\varphi} & (1+\cos\theta)\ee^{-\ii\varphi} & \sin\theta \cr
1-\cos\theta & -\sin\theta\ee^{\ii\varphi} & -\sin\theta\ee^{-\ii\varphi} & 1+\cos\theta 
\end{pmatrix}.
\end{equation}
\end{widetext}
Hence, one can introduce the probability amplitudes for the entangled spin states of created pairs,
\begin{align}\label{s5}
\mathcal{A}_S&=(\mathcal{A}_{\supdown}-\mathcal{A}_{\sdownup})/\sqrt{2}=(\mathcal{A}_{+,+}+\mathcal{A}_{-,-})/\sqrt{2},
\\
\mathcal{A}_+&=(\mathcal{A}_{\supup}+\mathcal{A}_{\sdowndown})/\sqrt{2}=(\mathcal{A}_{-,+}-\mathcal{A}_{+,-})/\sqrt{2},
\nonumber \\
\mathcal{A}_0&=(\mathcal{A}_{\supdown}+\mathcal{A}_{\sdownup})/\sqrt{2}=(\mathcal{A}_{+,+}-\mathcal{A}_{-,-})/\sqrt{2},
\nonumber \\
\mathcal{A}_-&=(\mathcal{A}_{\supup}-\mathcal{A}_{\sdowndown})/\sqrt{2}=-(\mathcal{A}_{+,-}+\mathcal{A}_{-,+})/\sqrt{2},
\nonumber 
\end{align}
such that the amplitude $\mathcal{A}_S$ is the scalar, i.e., is independent of the chosen spin quantization axis. Moreover, all these entangled 
amplitudes can be calculated from the amplitudes $\mathcal{A}^{(0)}_{\lambda_-,\lambda_+}$ by performing the matrix multiplication,
\begin{equation}\label{s6}
\begin{pmatrix}
\mathcal{A}_{S} & \mathcal{A}_{+} & \mathcal{A}_{0} & \mathcal{A}_{-} 
\end{pmatrix}^T
=\mathbb{B}
\begin{pmatrix}
\mathcal{A}^{(0)}_{+,+} & \mathcal{A}^{(0)}_{+,-} & \mathcal{A}^{(0)}_{-,+} & \mathcal{A}^{(0)}_{-,-} 
\end{pmatrix}^T.
\end{equation}
The superscript $T$ means the transposition and the matrix $\mathbb{B}$,
\begin{equation}\label{s7}
\mathbb{B}=\frac{1}{\sqrt{2}}
\begin{pmatrix}
1 & 0 & 0 & 1 \cr
0 & -\ee^{\ii\varphi} & \ee^{-\ii\varphi} & 0 \cr
\cos\theta & \sin\theta\ee^{\ii\varphi} & \sin\theta\ee^{-\ii\varphi} & -\cos\theta \cr
\sin\theta & -\cos\theta\ee^{\ii\varphi} & -\cos\theta\ee^{-\ii\varphi} & -\sin\theta 
\end{pmatrix},
\end{equation}
can be called the Bell matrix, as it transforms the amplitudes from the product spin states to the amplitudes for the maximally entangled spin states.

As the Bell matrix is unitary, i.e., $\mathbb{B}^{\dagger}\mathbb{B}=\mathbb{I}$, the spin summed up modulus squared amplitudes are conserved. Additionally, by writing down explicitly the angular dependence of the amplitudes and the $\mathbb{B}$ matrix, we arrive at
\begin{equation}\label{s9}
\begin{pmatrix}
\mathcal{A}_{S}(\theta',\varphi') \cr \mathcal{A}_{+}(\theta',\varphi') \cr \mathcal{A}_{0}(\theta',\varphi') \cr \mathcal{A}_{-}(\theta',\varphi') 
\end{pmatrix}
=\mathbb{B}(\theta',\varphi')[\mathbb{B}(\theta,\varphi)]^\dagger
\begin{pmatrix}
\mathcal{A}_{S}(\theta,\varphi) \cr \mathcal{A}_{+}(\theta,\varphi) \cr \mathcal{A}_{0}(\theta,\varphi) \cr \mathcal{A}_{-}(\theta,\varphi) 
\end{pmatrix},
\end{equation}
where $\mathbb{B}(\theta',\varphi')[\mathbb{B}(\theta,\varphi)]^\dagger$ has the block structure,
\begin{equation}\label{s10}
\mathbb{B}(\theta',\varphi')[\mathbb{B}(\theta,\varphi)]^\dagger=
\begin{pmatrix}
1 & \bm{0}^T \cr
\bm{0} & \mathbb{B}_V(\theta',\varphi';\theta,\varphi)
\end{pmatrix},
\end{equation}
with $\bm{0}^T=(0,0,0)$. This means that the amplitude $\mathcal{A}_{S}$, as the function of electron and positron momenta, is independent of 
the spin quantization axis. Note that
\begin{equation}\label{s11}
\begin{pmatrix}
\mathcal{A}_{S}(\theta,\varphi) \cr \mathcal{A}_{+}(\theta,\varphi) \cr \mathcal{A}_{0}(\theta,\varphi) \cr \mathcal{A}_{-}(\theta,\varphi) 
\end{pmatrix}
=\mathbb{B}(\theta,\varphi)[\mathbb{S}(\theta,\varphi)]^\dagger
\begin{pmatrix}
\mathcal{A}_{+,+}(\theta,\varphi) \cr\mathcal{A}_{+,-}(\theta,\varphi) \cr
\mathcal{A}_{-,+}(\theta,\varphi) \cr\mathcal{A}_{-,-}(\theta,\varphi)
\end{pmatrix},
\end{equation}
which is equivalent to the system of equations \eqref{s5}.

The method presented above applies to any electromagnetic field for which the potential vanishes asymptotically in the past and future. 
Cases for which the potential takes on different asymptotic values (as it happens, for instance, with the commonly used Sauter pulse) can also be 
described in this way. This, however, leads to additional notational complications, mainly due to the distinction between kinetic and dynamical 
momenta, which for asymptotic states means a shift by constant values. Therefore, in our theoretical studies as well as in the numerical analysis 
presented in Sec.~\ref{sec:num}, we limit ourselves to the case of vanishing potentials for both time infinities.

\subsection{Homogeneous electric field pulses}
\label{sec:timeF}

Numerical analysis of 
the Sauter-Schwinger effect in a time- and space-dependent electric field, even though intensively developed in recent theoretical studies, is still 
in its infancy. The reason being that strict ab initio calculations remain rather unfeasible, especially if the boundary-value problem has to be solved. 
This usually requires some approximations, the validity of which is not always clear. For this reason, in our numerical illustrations, we will 
consider the time-dependent electric pulses. Even in this case, various nonequivalent theoretical approaches exist in literature, the similarities 
and differences of which can be traced without further approximations. Such analysis can be called rigorous and might 
serve as a test of the applicability of various approximations, like the ones used in the worldline formalism. Therefore,  
we adapt the general method developed above to the case of homogeneous electric field pulses.

For homogeneous electric field pulses, electrons and positrons are generated with opposite momenta. As follows from the analysis presented in \cite{PhysRevD.110.116025}, in Eqs.~\eqref{fb1a} and \eqref{fb1b} one should make the substitutions,
\begin{equation}\label{tf1}
C^{(\beta)}_{\mathrm{F};\bm{p}_+,\lambda_+}(\bm{q},\lambda)=(2\pi)^3\delta^{(3)}(\bm{q}+\beta\bm{p}_+)
C^{(\beta)}_{\mathrm{F};\lambda_+}(\bm{q},\lambda).
\end{equation}
Thus, Eq.~\eqref{f15} can be rewritten in the form,
\begin{equation} \label{tf2}
\mathcal{A}(\bm{p}_-,\lambda_-;\bm{p}_+,\lambda_+)=(2\pi)^3\delta^{(3)}(\bm{p}_-+\bm{p}_+)\mathcal{A}^{(+)}_{\lambda_+}(\bm{p}_-,\lambda_-),
\end{equation}
with
\begin{equation} \label{tf2r}
\mathcal{A}^{(+)}_{\lambda_+}(\bm{p}_-,\lambda_-)=C^{(+)}_{\mathrm{F};\lambda_+}(\bm{p}_-,\lambda_-).
\end{equation}
Now, applying the standard rule, 
\begin{equation}\label{tf1a}
[(2\pi)^3\delta^{(3)}(\bm{p}_-+\bm{p}_+)]^2=V(2\pi)^3\delta^{(3)}(\bm{p}_-+\bm{p}_+), 
\end{equation}
where $V$ is the quantization volume, we arrive at
\begin{equation}\label{tf3}
\frac{|\mathcal{A}(\bm{p}_-,\lambda_-;\bm{p}_+,\lambda_+)|^2}{V}\! =\! (2\pi)^3\delta^{(3)}(\bm{p}_-+\bm{p}_+)f^{(+)}_{\lambda_+}(\bm{p}_-,\lambda_-),
\end{equation}
where
\begin{equation}\label{tf4}
f^{(+)}_{\lambda_+}(\bm{p}_-,\lambda_-)=|\mathcal{A}^{(+)}_{\lambda_+}(\bm{p}_-,\lambda_-)|^2
\end{equation}
is the probability distribution per unit volume for creation of electrons of momentum $\bm{p}_-$ and spin polarization $\lambda_-$ provided that positrons are generated with momentum $-\bm{p}_-$ and spin polarization $\lambda_+$. Moreover, by summing up the probability distributions over the positron spin polarizations, we obtain the reduced distributions,
\begin{equation}\label{tf4a}
f^{(+)}(\bm{p}_-,\lambda_-)=\sum_{\lambda_+=\pm}f^{(+)}_{\lambda_+}(\bm{p}_-,\lambda_-).
\end{equation}
Furthermore, since for particles the amplitudes for the spin polarization and projection are the same, we get the helicity-resolved momentum distributions for created electrons,
\begin{equation}\label{tf4b}
f_{\uparrow}^{(+)}(\bm{p}_-)=f^{(+)}(\bm{p}_-,+),\, 
f_{\downarrow}^{(+)}(\bm{p}_-)=f^{(+)}(\bm{p}_-,-),
\end{equation}
provided that the angles $\theta$ and $\varphi$ relate to the electron momentum $\bm{p}_-$.

Similarly to Eq.~\eqref{s5}, by using the short-hand notation,
\begin{equation}\label{tf5}
\mathcal{A}^{(+)}_{\lambda_-,\lambda_+}=\mathcal{A}^{(+)}_{\lambda_+}(\bm{p}_-,\lambda_-),
\end{equation}
one can introduce the spin-entangled probability amplitudes,
\begin{align}\label{s5a}
\mathcal{A}^{(+)}_S&=(\mathcal{A}^{(+)}_{\supdown}-\mathcal{A}^{(+)}_{\sdownup})/\sqrt{2}=(\mathcal{A}^{(+)}_{+,+}+\mathcal{A}^{(+)}_{-,-})/\sqrt{2},
\\
\mathcal{A}^{(+)}_+&=(\mathcal{A}^{(+)}_{\supup}+\mathcal{A}^{(+)}_{\sdowndown})/\sqrt{2}=(\mathcal{A}^{(+)}_{-,+}-\mathcal{A}^{(+)}_{+,-})/\sqrt{2},
\nonumber \\
\mathcal{A}^{(+)}_0&=(\mathcal{A}^{(+)}_{\supdown}+\mathcal{A}^{(+)}_{\sdownup})/\sqrt{2}=(\mathcal{A}^{(+)}_{+,+}-\mathcal{A}^{(+)}_{-,-})/\sqrt{2},
\nonumber \\
\mathcal{A}^{(+)}_-&=(\mathcal{A}^{(+)}_{\supup}-\mathcal{A}^{(+)}_{\sdowndown})/\sqrt{2}=-(\mathcal{A}^{(+)}_{+,-}+\mathcal{A}^{(+)}_{-,+})/\sqrt{2},
\nonumber 
\end{align}
such that the amplitude $\mathcal{A}^{(+)}_S$ is the scalar with respect to rotations of the spin quantization axis and the dependence of these amplitudes on the electron momentum $\bm{p}_-$ has been omitted. This allows us to define the momentum distribution of created electrons in the spin-entangled states,
\begin{align}\label{s5b}
(
f^{(+)}_{S}, f^{(+)}_{+}, 
f^{(+)}_{0}, f^{(+)}_{-}
)
=
(&
|\mathcal{A}^{(+)}_{S}|^2, |\mathcal{A}^{(+)}_{+}|^2, \\
& |\mathcal{A}^{(+)}_{0}|^2, |\mathcal{A}^{(+)}_{-}|^2
), \nonumber
\end{align}
that will be discussed below.

\section{Initial-value approach}
\label{sec:ret}

The analysis of the Feynman formalism carried out in Sec.~\ref{sec:spacetime} clearly shows that if we adopt the Feynman space-time approach 
to QED processes (or, equivalently, the scattering matrix theory), the problem of pair creation by a classical electromagnetic field from the vacuum 
can be reduced to solving the Dirac equation with Feynman or anti-Feynman boundary conditions. On the other hand, the frequently applied method 
of studying this process is to analyze solutions of the Dirac equation with normalized initial conditions, even though neither electrons nor positrons 
exist in the past. Regardless of this fact, one may wonder whether a modification of the Feynman formalism is possible, which unambiguously implies 
imposing initial conditions on solutions of the Dirac equation. For this purpose, let us replace in Eq.~\eqref{fff1} the Feynman propagator, 
$K_{\mathrm{F}}(y,x)$, by the retarded one, $K_{\mathrm{R}}(y,x)$. The latter fulfills the integral equations,
\begin{align}\label{r1}
\!\!\!\!  K_{\mathrm{R}}(y,x)\! =&S_{\mathrm{R}}(y\! -\! x)\! -\! \int\!\!  \dd^4zS_{\mathrm{R}}(y\! -\! z)e\slashed{A}(z)K_{\mathrm{R}}(z,x), \\
\! =&S_{\mathrm{R}}(y\! -\! x)\! -\!\!  \int\!\! \dd^4zK_{\mathrm{R}}(y,z)e\slashed{A}(z)S_{\mathrm{R}}(z\! -\! x), \nonumber
\end{align}
with the free-fermion retarded propagator defined by the integral,
\begin{equation}\label{r2}
S_{\mathrm{R}}(z)\! =\! -\! \int\! \frac{\dd^4q}{(2\pi)^4}\frac{\slashed{q}+mc}{(q^0\! +\! \ii\varepsilon)^2\! -\! \bm{q}^2\! -\! (mc)^2}\ee^{-\ii q\cdot z},  
\end{equation}
from which one derives 
\begin{equation}\label{r3}
\! \!  S_{\mathrm{R}}(y\! -\! x)\! =\! \ii\theta(y^0\! -\! x^0)\!\!  \sum_{\lambda,\beta=\pm}\! \int\! \frac{\dd^3q}{(2\pi)^3}
\chi^{(\beta)}_{\bm{q},\lambda}(y)\bar{\chi}^{(\beta)}_{\bm{q},\lambda}(x).
\end{equation}
Both, $S_{\mathrm{R}}(z)$ and $K_{\mathrm{R}}(y,x)$, satisfy the same inhomogeneous Dirac equations [Eqs.~\eqref{fff4} and~\eqref{fff4a}, respectively] 
as the Feynman propagators. The corresponding amplitude can be formulated similarly to Eqs.~\eqref{ff1} and~\eqref{ff1a},
\begin{widetext}
\begin{equation} \label{r4}
\mathcal{A}_{\mathrm{R}}(\bm{p}_{-},\lambda_{-}; \bm{p}_{\mathrm{i}},\lambda_{\mathrm{i}}) 
=\ii \lim_{y^0\rightarrow +\infty}\lim_{x^0\rightarrow -\infty}\int \dd^3x\dd^3y\bar{\chi}^{(+)}_{\bm{p}_{-},\lambda_{-}}(y)\gamma^0K_{\mathrm{R}}(y,x)\gamma^0\chi^{(-)}_{\bm{p}_{\mathrm{i}},\lambda_{\mathrm{i}}}(x),
\end{equation}
or, equivalently,
\begin{equation} \label{r5}
[\mathcal{A}_{\mathrm{R}}(\bm{p}_{-},\lambda_{-}; \bm{p}_{\mathrm{i}},\lambda_{\mathrm{i}})]^* 
=-\ii \lim_{y^0\rightarrow +\infty}\lim_{x^0\rightarrow -\infty}\int \dd^3x\dd^3y\bar{\chi}^{(-)}_{\bm{p}_{\mathrm{i}},\lambda_{\mathrm{i}}}(x)\gamma^0K_{\mathrm{A}}(x,y)\gamma^0\chi^{(+)}_{\bm{p}_{-},\lambda_{-}}(y),
\end{equation}
\end{widetext}
where $K_{\mathrm{A}}(x,y)=\gamma^0[K_{\mathrm{R}}(y,x)]^{\dagger}\gamma^0$ is the advanced propagator. However, the important difference arises in 
the limit $x^0\rightarrow -\infty$. Now, the negative energy solutions of the free-particle Dirac equation are defined in the past. Therefore, they 
do not correspond to positrons created in the future, as it is the case for the Feynman boundary conditions. For this reason, we have also replaced 
the labels $(\bm{p}_+,\lambda_+)$ by $(\bm{p}_{\mathrm{i}},\lambda_{\mathrm{i}})$. Moreover, if we assume that during the process the total electric 
charge is conserved then the modified Feynman formulas~\eqref{r4} and~\eqref{r5} indicate that the initial negative energy state 
$\chi^{(-)}_{\bm{p}_{\mathrm{i}},\lambda_{\mathrm{i}}}(x)$ describes electrons, as already noticed by Feynman (see, e.g.,~\cite{feynman1998quantum}). 
Note that Eq.~\eqref{r5} describes the final-value approach in which the final positive energy electron state is propagated backwards in time and 
projected on the initial negative energy electron state [see, e.g., Ref.~\cite{PhysRevD.94.065024}; for bosons, which are not considered here, 
such a possibility was already studied in~\cite{PhysRevD.2.1191}, cf. Eq.~(25)]. In this sense the initial- and final-value approaches are equivalent. 
As they are related by the complex conjugation, they are described in fact by the same Feynman-type formula. For this reason, in the following we will 
focus exclusively on the initial-value problem analysis.

The retarded amplitude, $\mathcal{A}_{\mathrm{R}}(\bm{p}_{-},\lambda_{-}; \bm{p}_{\mathrm{i}},\lambda_{\mathrm{i}})$, can be also put in the form, [cf., \eqref{f9}],
\begin{equation}\label{r6}
\mathcal{A}_{\mathrm{R}}(\bm{p}_-,\lambda_-;\bm{p}_\mathrm{i},\lambda_\mathrm{i})\! =\! -\ii\!\!\int\!  \dd^4y \bar{\chi}^{(+)}_{\bm{p}_{-},\lambda_{-}}\! (y)e\slashed{A}(y)\psi^{(-)}_{\mathrm{R};\bm{p}_\mathrm{i},\lambda_\mathrm{i}}\! (y),
\end{equation}
where
\begin{equation}\label{r7}
\psi^{(-)}_{\mathrm{R};\bm{p}_\mathrm{i},\lambda_\mathrm{i}}(y)=\chi^{(-)}_{\bm{p}_\mathrm{i},\lambda_\mathrm{i}}(y)-\! \int\! \dd^4x K_{\mathrm{R}}(y,x)e\slashed{A}(x)\chi^{(-)}_{\bm{p}_\mathrm{i},\lambda_\mathrm{i}}(x).
\end{equation}
This state fulfills the Dirac equation~\eqref{f10}, but in order to derive its asymptotic behavior, let us recall that similarly to~\eqref{f11},
\begin{equation}\label{r8}
K_{\mathrm{R}}(y,x)=\ii\theta(y^0-x^0)\sum_{\lambda,\beta=\pm}\int\frac{\dd^3q}{(2\pi)^3}
\chi^{(\beta)}_{\bm{q},\lambda}(y)\bar{\psi}^{(\beta)}_{\mathrm{A};\bm{q},\lambda}(x),
\end{equation}
for $y^0\rightarrow\pm\infty$ and fixed $x^0$. Here,
\begin{equation}\label{r9}
\psi^{(\beta)}_{\mathrm{A};\bm{q},\lambda}(x)=\chi^{(\beta)}_{\bm{q},\lambda}(x)
-\int\dd^4z K_{\mathrm{A}}(x,z)e\slashed{A}(z)\chi^{(\beta)}_{\bm{q},\lambda}(z),
\end{equation}
or
\begin{equation}\label{r10}
\bar{\psi}^{(\beta)}_{\mathrm{A};\bm{q},\lambda}(x)=\bar{\chi}^{(\beta)}_{\bm{q},\lambda}(x)
-\int\dd^4z \bar{\chi}^{(\beta)}_{\bm{q},\lambda}(z)e\slashed{A}(z)K_{\mathrm{R}}(z,x),
\end{equation}
is the advanced solution of the Dirac equation \eqref{f10}. Thus, for $y^0\rightarrow \infty$,
\begin{align}
\psi^{(-)}_{\mathrm{R};\bm{p}_\mathrm{i},\lambda_\mathrm{i}}(y)\! =&
\chi^{(-)}_{\bm{p}_\mathrm{i},\lambda_\mathrm{i}}(y)
 \label{r11} \\
+&\! \sum_{\beta,\lambda=\pm}\! \int\! \frac{\dd^3q}{(2\pi)^3}C^{(\beta)}_{\mathrm{R};\bm{p}_\mathrm{i},\lambda_\mathrm{i}}(\bm{q},\lambda)\chi^{(\beta)}_{\bm{q},\lambda}(y),
\nonumber 
\end{align}
and for $y^0\rightarrow -\infty$,
\begin{equation}
\psi^{(-)}_{\mathrm{R};\bm{p}_\mathrm{i},\lambda_\mathrm{i}}(y)\! =
\chi^{(-)}_{\bm{p}_\mathrm{i},\lambda_\mathrm{i}}(y)
, \label{r12}
\end{equation}
where
\begin{equation}\label{r13}
C^{(\beta)}_{\mathrm{R};\bm{p}_\mathrm{i},\lambda_\mathrm{i}}(\bm{q},\lambda)=-\ii\int\dd^4x \bar{\psi}^{(\beta)}_{\mathrm{A};\bm{q},\lambda}(x)e\slashed{A}(x)\chi^{(-)}_{\bm{p}_\mathrm{i},\lambda_\mathrm{i}}(x).
\end{equation}
Now, the negative energy solutions are initially fixed and as the result of time-evolution one gets the negative and positive energy wave packets in the future. In particular [cf., \eqref{f15}],
\begin{equation}\label{r14}
\mathcal{A}_{\mathrm{R}}(\bm{p}_-,\lambda_-;\bm{p}_\mathrm{i},\lambda_\mathrm{i})=C^{(+)}_{\mathrm{R};\bm{p}_\mathrm{i},\lambda_\mathrm{i}}(\bm{p}_-,\lambda_-).
\end{equation}
However, in addition to the spin and momentum amplitude for the created electrons, we also obtain the corresponding amplitude for the created particles 
described by the negative energy states,
\begin{align}\label{r14a}
\tilde{\mathcal{A}}_{\mathrm{R}}(\bm{p}_+,\lambda_+;\bm{p}_\mathrm{i},\lambda_\mathrm{i})&=(2\pi)^3\delta^{(3)}(\bm{p}_+-\bm{p}_{\mathrm{i}})\delta_{\lambda_\mathrm{i},\lambda_+} \\
&+C^{(-)}_{\mathrm{R};\bm{p}_\mathrm{i},\lambda_\mathrm{i}}(\bm{p}_+,\lambda_+). \nonumber
\end{align}
The amplitude~\eqref{r14a} also follows from the modified Feynman formula~\eqref{r4} in which the positive energy state, 
$\chi^{(+)}_{\bm{p}_{-},\lambda_{-}}$, is replaced by the negative energy state, $\chi^{(-)}_{\bm{p}_{+},\lambda_{+}}$, and describes the scattering process of negative energy electrons in the Dirac sea. Note that such a replacement 
is forbidden in the original Feynman formula~\eqref{ff1} due to the Feynman boundary conditions, as it follows for instance from~\eqref{f11}. Since 
the time evolution defined by the Dirac equation is unitary, both positive and negative energy solutions have to describe electrons (as already noticed 
by Feynman), if one requires the conservation of the electric charge. Finally, similarly to the Feynman and anti-Feynman states, 
Eqs.~\eqref{r6},~\eqref{r13} and~\eqref{r14} resemble the scattering amplitude expressed by the scattering states with either the outgoing spherical 
waves (in the present case the retarded solutions) or the incoming spherical waves (here, the advanced solutions).

The above discussion has led us to the solution of the Dirac equation with the initial condition imposed on the negative energy part of the Dirac 
bispinor, from which the positive energy part in the future is determined. But the similar analysis can be carried out for the case when the initial 
condition is defined by the positive energy part of the Dirac bispinor, whereas its negative energy part is evaluated. In order to achieve this goal, 
one should replace the Feynman- by the advanced propagator, and perform the identical analysis for the complex conjugate amplitude.

\subsection{Homogeneous electric field pulses}
\label{sec:timeR}

Following the same analysis as in Sec.~\ref{sec:timeF}, for the initial-value problem and for homogeneous electric field pulses, the amplitudes~\eqref{r14} and~\eqref{r14a} take the forms,
\begin{equation}\label{tr1}
\mathcal{A}_{\mathrm{R}}(\bm{p}_-,\lambda_-;\bm{p}_\mathrm{i},\lambda_\mathrm{i})=(2\pi)^3\delta^{(3)}(\bm{p}_-+\bm{p}_{\mathrm{i}})\mathcal{A}^{(+)}_{\mathrm{R};\lambda_\mathrm{i}}(\bm{p}_-,\lambda_-)
\end{equation}
and
\begin{align}\label{tr1a}
\tilde{\mathcal{A}}_{\mathrm{R}}(\bm{p}_+,\lambda_+;\bm{p}_\mathrm{i},\lambda_\mathrm{i})=(2\pi)^3\delta^{(3)}(\bm{p}_+-\bm{p}_{\mathrm{i}})\tilde{\mathcal{A}}^{(-)}_{\mathrm{R};\lambda_\mathrm{i}}(\bm{p}_+,\lambda_+) .
\end{align}
Hence, we obtain the probability distributions per unit volume for created electrons of the momentum $\bm{p}_-$ and spin polarization $\lambda_-$,
\begin{equation}\label{tr2}
f^{(+)}_{\mathrm{R};\lambda_\mathrm{i}}(\bm{p}_-,\lambda_-)=|\mathcal{A}^{(+)}_{\mathrm{R};\lambda_\mathrm{i}}(\bm{p}_-,\lambda_-)|^2,
\end{equation}
provided that the negative energy initial state is defined by the momentum $\bm{p}_\mathrm{i}=-\bm{p}_-$ and spin polarization $\lambda_\mathrm{i}$. 
Moreover, for the momentum and spin-resolved electron distribution we obtain
\begin{equation}\label{tr3}
f^{(+)}_{\mathrm{R}}(\bm{p}_-,\lambda_-)=\sum_{\lambda_{\mathrm{i}}=\pm}f^{(+)}_{\mathrm{R};\lambda_\mathrm{i}}(\bm{p}_-,\lambda_-).
\end{equation}
The above distributions for created electrons have, however, a certain drawback; they use the spin polarization of hypothetical positrons that do not exist in the past.

\subsection{Spin-resolved amplitudes}
\label{sec:spininitialvalue}

Similarly to the case analyzed in Sec.~\ref{sec:spin}, one can define the spin-resolved amplitudes for the initial-value approach. However, now we 
have two copies of the spin-resolved momentum distributions of created pairs. First, when in the past the negative energy state is fixed and one 
determines the spin-resolved amplitudes for the created electrons, as it is the case studied in our analysis. Second, when the spin-resolved amplitudes 
for produced positrons are considered provided that the positive energy state is fixed in the past. Considering the first case, let us introduce 
the abbreviation [cf., Eq.~\eqref{tr1}],
\begin{equation}\label{si1}
\mathcal{A}^{(+)}_{\mathrm{R};\lambda_-,\lambda_{\mathrm{i}}}=\mathcal{A}^{(+)}_{\mathrm{R};\lambda_\mathrm{i}}(\bm{p}_-,\lambda_-).
\end{equation}
We also denote as $\mathcal{A}^{(+)(0)}_{\mathrm{R};\lambda_-,\lambda_{\mathrm{i}}}$ the amplitude 
$\mathcal{A}^{(+)}_{\mathrm{R};\lambda_-,\lambda_{\mathrm{i}}}$ calculated for the spin quantization axis defined by $\theta=\varphi=0$. 
Then, following the same discussion as in Sec.~\ref{sec:spin}, we calculate the amplitudes for the spin quantization axis defined by an arbitrary polar 
$\theta$ and azimuthal $\varphi$ angles. For instance, for the negative energy states in the past and for 
the positive energy states in the future representing electrons, we have
\begin{equation}\label{si2}
\begin{pmatrix}
\mathcal{A}^{(+)}_{\mathrm{R};+,+} \cr\mathcal{A}^{(+)}_{\mathrm{R};+,-} \cr
\mathcal{A}^{(+)}_{\mathrm{R};-,+} \cr\mathcal{A}^{(+)}_{\mathrm{R};-,-}
\end{pmatrix}
=\mathbb{S}(\theta,\varphi)
\begin{pmatrix}
\mathcal{A}^{(+)(0)}_{\mathrm{R};+,+} \cr\mathcal{A}^{(+)(0)}_{\mathrm{R};+,-} \cr
\mathcal{A}^{(+)(0)}_{\mathrm{R};-,+} \cr\mathcal{A}^{(+)(0)}_{\mathrm{R};-,-}
\end{pmatrix}
.
\end{equation}
As the matrix $\mathbb{S}$ defined in Eq.~\eqref{ss2} is unitary, the spin summed up probabilities are conserved. Moreover, since for particles 
the amplitudes for the spin polarization and spin projection are the same, the helicity-resolved momentum distributions for created electrons 
[cf., Eq.~\eqref{tr3}] are 
\begin{equation}\label{si3}
f_{\mathrm{R};\uparrow}^{(+)}(\bm{p}_-)=f_{\mathrm{R}}^{(+)}(\bm{p}_-,+),\, 
f_{\mathrm{R};\downarrow}^{(+)}(\bm{p}_-)=f_{\mathrm{R}}^{(+)}(\bm{p}_-,-),
\end{equation}
with the angles $\theta$ and $\varphi$ related to the electron momentum $\bm{p}_-$.

Similarly to the boundary-value approach one can also introduce the spin-entangled amplitudes. However, since for the initial-value approach 
the amplitudes~\eqref{si2} are defined for spins in the past for the negative energy states and in the future for the created electrons, 
we presume that such entanglement could be of little interest for quantum information theories. Nevertheless, we will consider this topic, 
as the spin entanglement for the initial-value problem has been already studied 
in the context of dynamical Sauter-Schwinger effect~\cite{PhysRevD.95.036006,PhysRevA.99.032340}. To this end, we define the 
spin-entangled amplitudes for the initial-value approach,
\begin{equation}\label{si2a}
\begin{pmatrix}
\mathcal{A}^{(+)}_{\mathrm{R};S} \cr\mathcal{A}^{(+)}_{\mathrm{R};+} \cr
\mathcal{A}^{(+)}_{\mathrm{R};0} \cr\mathcal{A}^{(+)}_{\mathrm{R};-}
\end{pmatrix}
=\mathbb{B}
\begin{pmatrix}
\mathcal{A}^{(+)(0)}_{\mathrm{R};+,+} \cr\mathcal{A}^{(+)(0)}_{\mathrm{R};+,-} \cr
\mathcal{A}^{(+)(0)}_{\mathrm{R};-,+} \cr\mathcal{A}^{(+)(0)}_{\mathrm{R};-,-}
\end{pmatrix}.
\end{equation}
As for the boundary-value spin-entangled amplitude, also $\mathcal{A}^{(+)}_{\mathrm{R};S}$ is independent of the angles defining the spin quantization 
axis. This also allows us to define the momentum distribution of created electrons in the spin-entangled states,
\begin{align}\label{si2b}
(
f^{(+)}_{\mathrm{R};S}, f^{(+)}_{\mathrm{R};+}, 
f^{(+)}_{\mathrm{R};0}, f^{(+)}_{\mathrm{R};-}
)
=
(&
|\mathcal{A}^{(+)}_{\mathrm{R};S}|^2, |\mathcal{A}^{(+)}_{\mathrm{R};+}|^2, \\
&|\mathcal{A}^{(+)}_{\mathrm{R};0}|^2, |\mathcal{A}^{(+)}_{\mathrm{R};-}|^2
), \nonumber
\end{align}
that can be compared with the corresponding distributions defined by Eq.~\eqref{s5b}.

The approach based on solutions of the Dirac equation with an initial condition is commonly used in studies of the dynamical Sauter-Schwinger process 
(see, e.g.,~\cite{avetissian2002pair,blinne2014pair,blinne2016wigner,doi:10.1080/00107510903450559,li2023phasepair,xylm-fnt7,PhysRevD.111.076017,PhysRevD.110.076028,PhysRevD.110.056013,PhysRevD.111.056020,brass2025relative}). 
However, as we have demonstrated above, this method follows from the modified Feynman formalism, consisting in replacing the Feynman- by the retarded propagator. 
Still, it led, for instance, to the development of QKE~\cite{PhysRevD.110.L011901,PhysRevResearch.6.043009,PhysRevD.98.056009,PhysRevA.100.012104,PhysRevA.100.062116,Sah_Singh_2025,DeepakSah2026} and DHW~\cite{bialynicki1991diracvacuum,1wms-rjwl,mgst-vn3c}
methods, which were frequently applied for $e^-e^+$ pair creation by time-dependent electric fields. In this case, it was also proven that 
QKE and DHW methods are equivalent~\cite{Bechler_Krajewska_CajiaoVelez_Kaminski_2023}. However, one can wonder under which circumstances both approaches, i.e., 
based on the intial- and on the boundary conditions, provide similar results.
To address this problem, we present in Sec.~\ref{sec:qke} one of possible variants of the kinetic equations formalism. Then, 
in Sec.~\ref{sec:num}, we compare its numerical results with the results arising from solving the corresponding boundary-value problem.

\subsection{Quantum Kinetic Equations}
\label{sec:qke}

Let us shortly describe the QKE approach~\cite{PhysRevD.110.L011901,PhysRevResearch.6.043009}. To this end we assume that the vector potential depends 
only on time and the electric field strength equals $\bm{\mathcal{E}}(t)=-\dot{\bm{A}}(t)$, where overdot means the time derivative. In such a situation, 
if the created electrons in the future have the momentum $\bm{p}$, then negative energy solutions in the past and future have the momentum $-\bm{p}$. 
This means that all distributions depend only on one momentum. Let us introduce the time-dependent four-momentum $p(t)=(p^0(t),\bm{p}-e\bm{A}(t))$ on 
the electron mass shell (i.e., $[p(t)]^2=\me^2c^2$) and define (following Refs.~\cite{PhysRevD.110.L011901,PhysRevResearch.6.043009}) the functions,
\begin{align}\label{qke3}
\bm{\mu}_1(t)&=e\frac{\bm{p}(t)\times \bm{\mathcal{E}}(t)}{2p^0(t)(p^0(t)+\me c)}, 
\\
\bm{\mu}_2(t)&=e\frac{[\bm{p}(t)\cdot\bm{\mathcal{E}}(t)]\bm{p}(t)/p^0(t)-(p^0(t)+\me c)\bm{\mathcal{E}}(t)}{2p^0(t)(p^0(t)+\me c)}, \nonumber 
\end{align}
and
\begin{align}\label{qke2}
\bm{f}_j(p(t),t)&=\bm{\mu}_j(t)\times \bm{f}(p(t),t), \\
\bm{u}_j(p(t),t)&=\bm{\mu}_j(t)\times \bm{u}(p(t),t),\nonumber \\
\bm{v}_j(p(t),t)&=\bm{\mu}_j(t)\times \bm{v}(p(t),t) , \nonumber 
\end{align}
for $j=1,2$. Then the kinetic equations adopt the form,
\begin{align}\label{qke1}
\dot{f}(p(t),t)&=-2\bm{\mu}_2(t)\cdot\bm{u}(p(t),t), \\
\dot{\bm{f}}(p(t),t)&=2\bm{f}_1(p(t),t)-2\bm{v}_2(p(t),t))   , \nonumber \\
\dot{\bm{u}}(p(t),t)&=2\bm{u}_1(p(t),t)+(2f(p(t),t)-1)\bm{\mu}_2(t) \nonumber \\
                    &+2cp^0(t)\bm{v}(p(t),t)    , \nonumber \\
\dot{\bm{v}}(p(t),t)&=2\bm{v}_1(p(t),t)-2\bm{f}_2(p(t),t)-2cp^0(t)\bm{u}(p(t),t)    . \nonumber  
\end{align}
This system of equations has to be solved with the initial conditions such that the dimensionless functions $f$, $\bm{f}$, $\bm{u}$, and $\bm{v}$, 
as well as the electric field $\bm{\mathcal{E}}(t)$, vanish in the past. For our further purpose two of these distributions in the future will be used (the superscript $(+)$ means that these are the distributions for electrons, i.e., for particles),
\begin{equation}\label{qke4}
\! f_\mathrm{K}^{(+)}(\bm{p})=\!\! \lim_{t\rightarrow +\infty}\!\! f(p(t),t), \,
\bm{f}_\mathrm{K}^{(+)}(\bm{p})=\!\! \lim_{t\rightarrow +\infty}\!\! \bm{f}(p(t),t). 
\end{equation}
According to Refs.~\cite{PhysRevD.110.L011901,PhysRevResearch.6.043009}, their physical interpretation is such that $2f_\mathrm{K}^{(+)}(\bm{p})$ is the number of created electrons per unit volume of momentum $\bm{p}$, whereas the corresponding densities of created electrons for different helicities are equal to
\begin{equation}\label{qke5a}
 f_{\mathrm{K};\uparrow}^{(+)}(\bm{p})=f_\mathrm{K}^{(+)}(\bm{p})+\frac{\bm{p}\cdot\bm{f}_\mathrm{K}^{(+)}(\bm{p})}{|\bm{p}|}
\end{equation}
and
\begin{equation}\label{qke5b}
f_{\mathrm{K};\downarrow}^{(+)}(\bm{p})=f_\mathrm{K}^{(+)}(\bm{p})-\frac{\bm{p}\cdot\bm{f}_\mathrm{K}^{(+)}(\bm{p})}{|\bm{p}|}. 
\end{equation}
On the other hand, the QKE approach is equivalent to the kinetic equations that follow from the DHW formalism~\cite{bialynicki1991diracvacuum}, 
as discussed in~\cite{PhysRevResearch.6.043009}. Additionally, one can show analytically (see, e.g., \cite{Bechler_Krajewska_CajiaoVelez_Kaminski_2023}) and check 
numerically (see, e.g., \cite{PhysRevD.110.116025}) that the kinetic equations can be derived directly from the Dirac equation with the initial 
conditions imposed on its solutions. Hence, the QKE method is also equivalent to the modified space-time Feynman approach in which the Feynman 
propagator $S_{\mathrm{F}}$ is replaced by either the retarded propagator $S_{\mathrm{R}}$, used when negative energy solutions are treated as the initial 
state to determine the momentum distribution of created electrons, or by the advanced propagator $S_{\mathrm{A}}$, used when positive energy solutions 
serve as the initial condition to calculate the distributions of created negative energy electrons (interpreted as positrons).
Consequently, these momentum distributions represent the transition probabilities per unit volume from initial negative energy states in the Dirac sea to final positive energy states of created electrons with momentum
${\bm p}$. This implies that the equation,
\begin{equation}\label{qke6}
 2f_\mathrm{K}^{(+)}(\bm{p})=\!\! \sum_{\lambda_{\mathrm{i}},\lambda_-=\pm} f^{(+)}_{\mathrm{R};\lambda_{\mathrm{i}}}(\bm{p},\lambda_-),
\end{equation}
has to be satisfied. Since this is the exact analytical result, it can serve as a test for our numerical studies. It appears that this equation 
can be fulfilled with the error smaller than $10^{-13}$ in the double precision by applying the standard Dormand-Prince method~\cite{DOPRI,Hairer}. 
However, in order to speed up numerical calculations, we will chose the accuracy of $10^{-10}$. Most importantly, it is 
the interpretation of the helicity-resolved distributions, Eqs.~\eqref{qke5a} and~\eqref{qke5b}, that will be the focus of our numerical analysis
in Sec.~\ref{sec:num}.

\section{Numerical analysis}
\label{sec:num}

Our numerical analysis compares predictions from the boundary- and initial-value approaches using a spatially homogeneous electric field pulse 
with a peak strength and frequency below $\mathcal{E}_{S}$ and $m_{\rm e}c^{2}$, respectively. As noted in Ref.~\cite{PhysRevD.110.116025}, for such 
electric field paramaters, both methods 
yield nearly identical spin-summed momentum distributions of created pairs. However, unlike QKE and DHW formalisms, solving the Dirac equation provides complete 
dynamical information, including all spin-resolved complex probability amplitudes. Following proposals presented in 
Refs.~\cite{PhysRevLett.104.220404,avetissian2002pair,kohlfurst2022collidinglaser}, we model the electric field pulse as the superposition of at least 
six laser pulses in a region where their magnetic components cancel out.

\begin{figure}
\includegraphics[width=7.5cm]{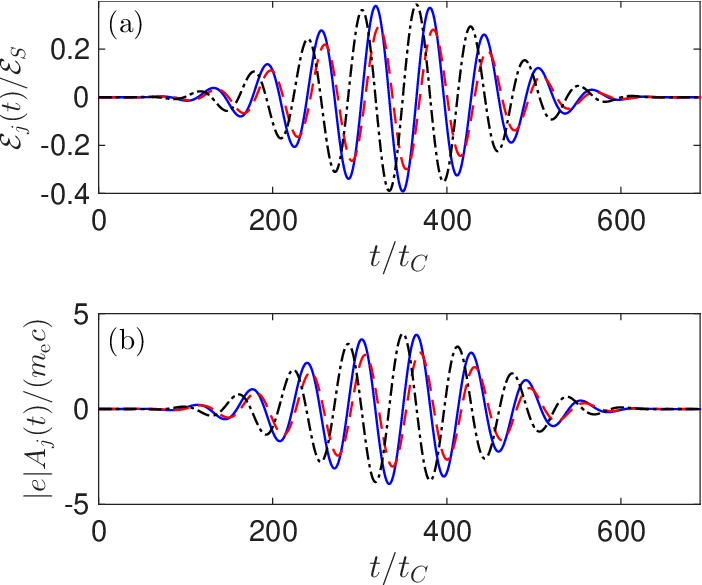}
\caption{(a) Components of the electric field pulse~\eqref{num5} for $j=1$ (solid line), $j=2$ (dashed line), and $j=3$ (dot-dashed line).
The electric field parameters has been chosen as: $\mathcal{E}_0=0.3\mathcal{E}_S$, $\omega=0.1m_\mathrm{e}c^2$, $N_\mathrm{osc}=11$, $\delta_1=-\pi/4$, 
$\delta_2=\delta_3=\pi/4$, $\chi_1=\chi_3=0$, $\chi_2=\pi/4$. (b) Corresponding components of the vector potential, defined by Eqs.~\eqref{num4}
and~\eqref{num5}.
}
\label{fpola}
\end{figure}

In order to define the electric field pulse, we introduce three unit and orthogonal vectors $\bm{e}_j$, $j=1,2,3$, such that $\bm{e}_1\times\bm{e}_2=\bm{e}_3$, and define the vector functions,
\begin{align}\label{num1}
\bm{\mathcal{E}}_{[k,\ell,j]}(t)=\mathcal{E}_0 F(t)&
\bigl(\cos(\omega t+\chi_j)\cos\delta_j\bm{e}_k \\
& +\sin(\omega t+\chi_j)\sin\delta_j\bm{e}_\ell\bigr). \nonumber
\end{align}
Here, the multi-index $[k,\ell,j]$ is a permutation of $[1,2,3]$, whereas $\mathcal{E}_0$, $\omega$, and $\chi_j$ are, respectively, the amplitude, 
central frequency and carrier envelope phase, and $\delta_j$ determines the field polarization. Furthermore,
\begin{equation}\label{num2}
F(t)=\begin{cases}
\sin^4\Bigl(\frac{\omega t}{2N_\mathrm{osc}}\Bigr), & 0<\omega t<2\pi N_\mathrm{osc}, \cr
0, & \textrm{otherwise},
\end{cases}
\end{equation}
is the envelope with $N_\mathrm{osc}$ cycles in the pulse. These functions define the electric field pulse,
\begin{equation}\label{num3}
\bm{\mathcal{E}}(t)=\bm{\mathcal{E}}_{[1,2,3]}(t)+\bm{\mathcal{E}}_{[2,3,1]}(t)+\bm{\mathcal{E}}_{[3,1,2]}(t),
\end{equation}
which will be used in our analysis. Moreover, the vector potential is equal to
\begin{equation}\label{num4}
\bm{A}(t)=-\int_0^t \dd\tau \bm{\mathcal{E}}(\tau),
\end{equation}
which, together with the electric field, vanishes for $t<0$ and $t>2\pi N_\mathrm{osc}$ provided that $N_\mathrm{osc}>2$. The example of such a pulse 
is presented in Fig.~\ref{fpola} for the components of the electric field and the vector potential,
\begin{equation}\label{num5}
\mathcal{E}_j(t)=\bm{e}_j\cdot\bm{\mathcal{E}}(t), \quad A_j(t)=\bm{e}_j\cdot\bm{A}(t).
\end{equation}
Note that the electric field parameters specified in Fig.~\ref{fpola} are used in the following numerical illustrations, unless otherwise stated. Even though the vectors 
$\bm{e}_j$ can be arbitrary oriented in space, we choose them as the Cartesian vectors such that $\bm{e}_1=\bm{e}_x$, $\bm{e}_2=\bm{e}_y$ and $\bm{e}_3=\bm{e}_z$.

\subsection{QKE vs. initial-value approach}\label{sec:QKEversusInitial}

\begin{figure}
\includegraphics[width=7.5cm]{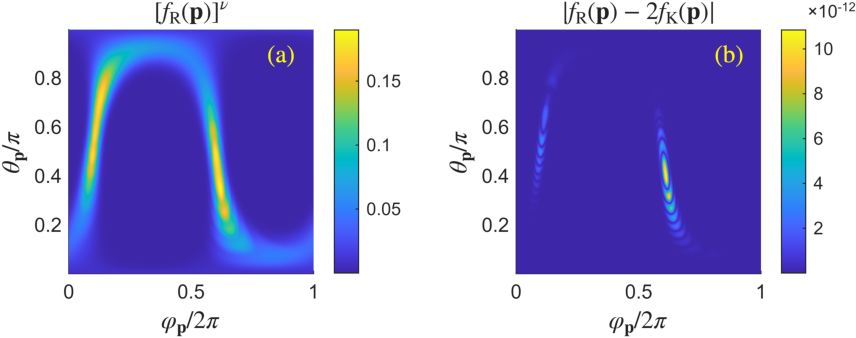}
\caption{(a) Color mappings of the spin summed up momentum distribution of electrons with momenta~\eqref{qqke1}, calculated
via the initial-value approach. Here, $f_\mathrm{R}(\bm{p})$ is raised to power $\nu=1/4$ for visual purposes. Discrepancies with the QKE theory 
are shown in panel (b). Since they are smaller than the accuracy of the numerical analysis (set to $10^{-10}$), we conclude that both methods generate identical results.
}
\label{figCompareRKexK}
\end{figure}

The saddle point method, pioneered by Keldysh in the analysis of ionization~\cite{Keldysh1964} and later developed in studies of other quantum 
processes, including the $e^-e^+$ pair creation~\cite{PhysRevD.2.1191,Popov2002,PhysRevLett.104.220404}, estimates positions of maxima 
in the corresponding momentum distributions. Qualitatively, it indicates that maxima in created electron distributions occur at momenta near 
the extreme values of $e\bm{A}(t)$. For the considered electric pulse (Fig.~\ref{fpola}), these maxima can be expected in the vicinity of 
$|\bm{p}|=5\me c$. For this reason, Fig.~\ref{figCompareRKexK}(a) shows the spin summed up angular distribution of created electrons for momenta
\begin{align}\label{qqke1}
\bm{p}(\theta_{\bm{p}},\varphi_{\bm{p}})=5\me c&(\sin\theta_{\bm{p}} \cos\varphi_{\bm{p}} \bm{e}_1 \\
&+\sin\theta_{\bm{p}} \sin\varphi_{\bm{p}} \bm{e}_2+\cos\theta_{\bm{p}} \bm{e}_3), \nonumber
\end{align}
calculated with the use of the initial-value approach. In Fig.~\ref{figCompareRKexK}(b) we present the discrepancies between the results obtained 
from the QKE and initial-value formalisms. Note that their maximum value is smaller then the accuracy of numerical analysis, settled to $10^{-10}$. 
For this reason, we claim that both the QKE and initial-value approaches generate identical results. This is expected, as QKE can be analytically 
derived from the Dirac equation, as shown for the equivalent DHW formalism (e.g., see~\cite{Bechler_Krajewska_CajiaoVelez_Kaminski_2023}). While some discrepancies between 
these approaches have been predicted in Ref.~\cite{PhysRevD.111.056020}, we attribute them to approximations used in the Dirac equation analysis, such as 
those noted below Eq.~(10) in that work. Furthermore, contrary to claims in, full electron-positron helicity-resolved momentum distributions have 
been analyzed in Ref.~\cite{PhysRevD.110.116025} without additional approximations. This contrasts with other studies~\cite{PhysRevD.110.L011901,PhysRevResearch.6.043009}, 
which evaluated only the electron helicity-resolved distributions.

\begin{figure}
\includegraphics[width=7.5cm]{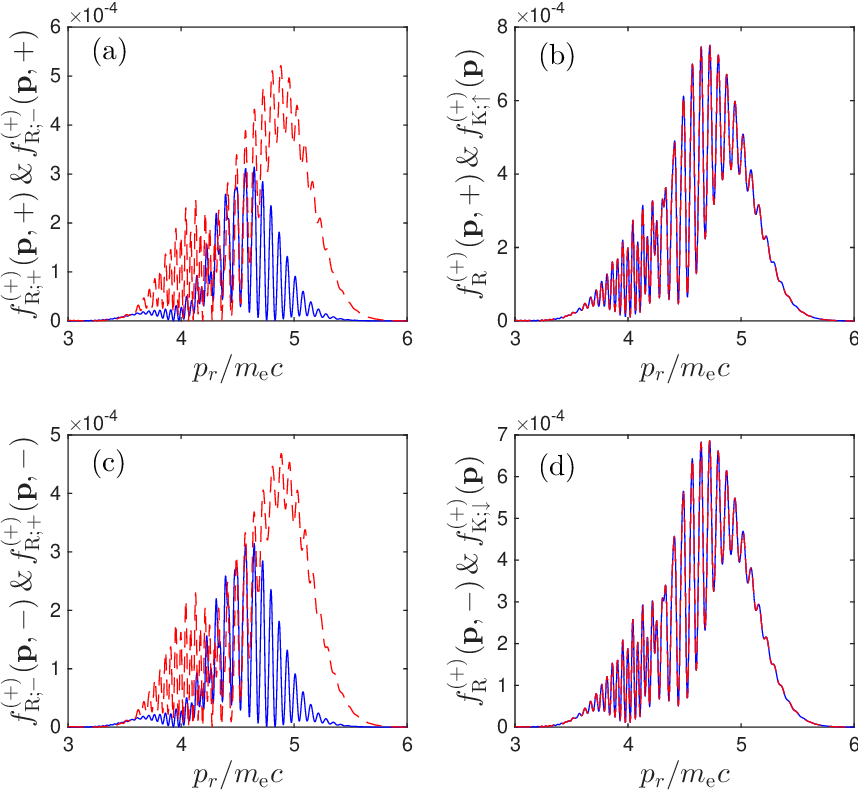}
\caption{Panels (a) and (c) show the spin-resolved momentum distributions [Eq.~\eqref{tr2}] for $\bm{p}$ defined by Eq.~\eqref{qqke2} and 
for $\varphi_{\bm{p}}=\pi/5$. The solid lines are for the same whereas the dashed lines are for different spin polarizations. 
Panels (b) and (d) compare the results for helicity-resolved distributions calculated with the initial-value approach [solid line for the distributions 
defined by Eqs.~\eqref{tr3} or~\eqref{si3}] and the QKE theory [dashed lines for the distributions~\eqref{qke5a} or~\eqref{qke5b}]. Perfect agreement 
of the results that follow from these two methods is observed, with the error smaller than $10^{-10}$.
}
\label{figCompareQKEUpDown}
\end{figure}

In order to investigate the helicity-resolved momentum distributions, we choose a particular direction in space defined by the polar, 
$\theta_{\bm{p}}=\pi/2$, and azimuthal, $\varphi_{\bm{p}}=\pi/5$, angles. This choice is motivated by 
the fact that for these angles and for $|\bm{p}|=5\me c$, the momentum distribution reaches a significant value [cf., Fig.~\ref{figCompareRKexK}(a)]. 
We also define the momentum,
\begin{equation}\label{qqke2}
\bm{p}(p_r,\varphi_{\bm{p}})=p_r(\cos\varphi_{\bm{p}} \bm{e}_1+\sin\varphi_{\bm{p}} \bm{e}_2),
\end{equation}
for fixed $\varphi_{\bm{p}}=\pi/5$, and investigate the dependence of the helicity-resolved distributions while changing $p_r$. The results are 
presented in Fig.~\ref{figCompareQKEUpDown} for $3\me c<p_r<6\me c$, for which the distributions acquire significant values. In 
Figs.~\ref{figCompareQKEUpDown}(a) and~\ref{figCompareQKEUpDown}(c), four momentum distributions 
$f^{(+)}_{\mathrm{R};\lambda_\mathrm{i}}(\bm{p},\lambda)$ are shown, where $\lambda$ labels two helicities of created electrons and 
$\lambda_\mathrm{i}$ denotes the spin degrees of freedom for the initial negative energy states. The electron helicity-resolved distributions 
[cf., Eq.~\eqref{tr3}] are presented in Figs.~\ref{figCompareQKEUpDown}(b) and~\ref{figCompareQKEUpDown}(d), together with the corresponding distributions \eqref{qke5a} and \eqref{qke5b} that follow from QKE. Discrepancies between these distributions are smaller than the precision of our numerical analysis, which means that for the initial spin summed up distributions both the initial-value and QKE methods reproduce the same results.

In closing this section, we conclude that QKE approach provides no unique physical insights into pair creation beyond what can
be achieved using the Dirac equation with normalized initial (or final) conditions. For this reason, we compare below only the initial- 
and boundary-value approaches.

\begin{figure}
\includegraphics[width=7.5cm]{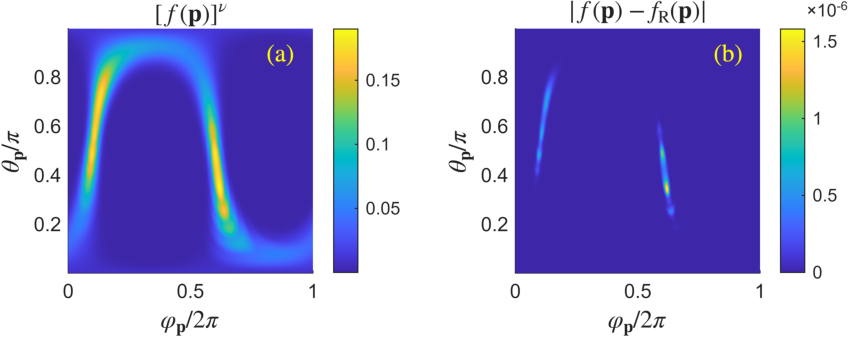}
\caption{(a) Color mappings of the spin summed up momentum distribution of electrons for momenta defined by Eq.~\eqref{qqke1}, calculated with
the boundary-value approach. For visual purposes, $f(\bm{p})$ is raised to power $\nu=1/4$. Discrepancies with the initial-value theory are shown 
in panel (b). Since they are by four orders of magnitude larger than the accuracy of our numerical analysis (equal to $10^{-10}$), we conclude that both methods 
generate essentially different results.
}
\label{figCompareBRexK}
\end{figure}

\subsection{Boundary- vs. initial-value approaches}\label{sec:BoundaryversusInitial}

\begin{figure}
\includegraphics[width=7.5cm]{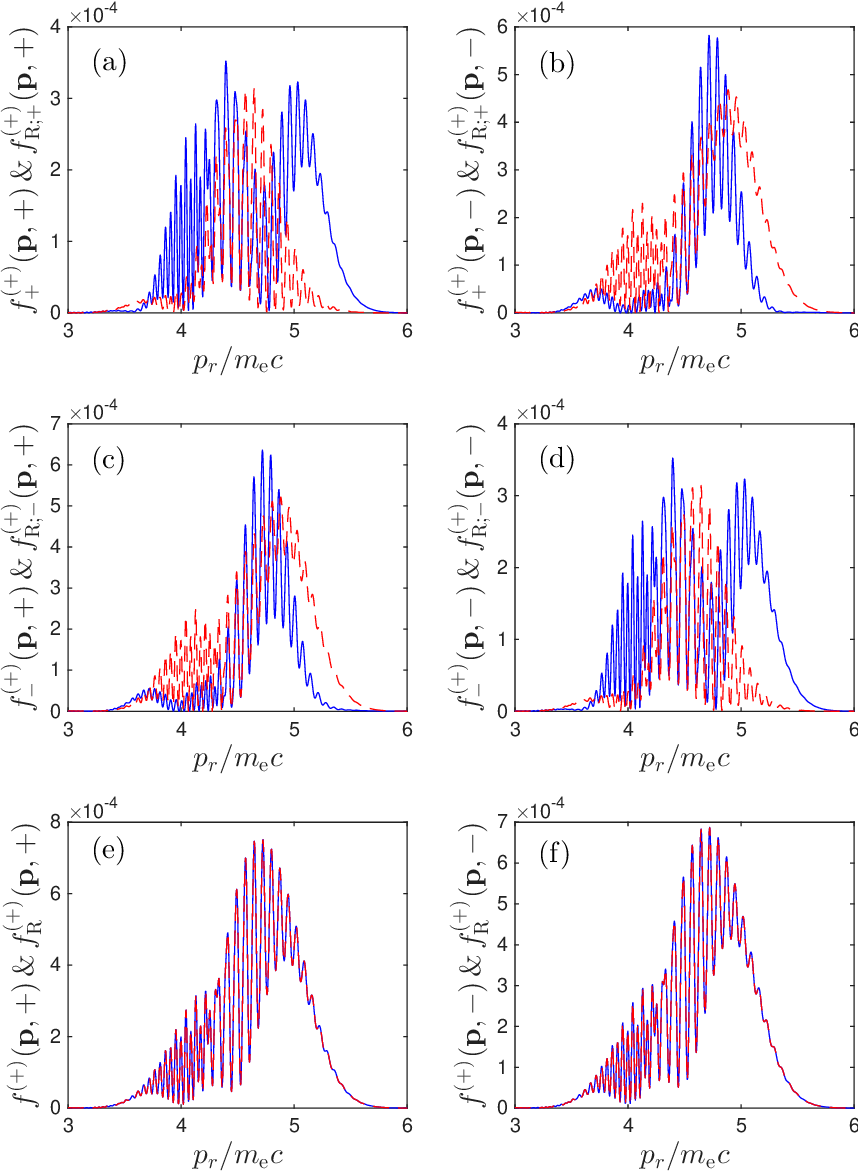}
\caption{Panels (a)-(d) show four spin-resolved distributions for $\bm{p}$ defined in Eq.~\eqref{qqke2} and for $\varphi_{\bm{p}}=\pi/5$, calculated
with either the boundary-value- (solid lines) or the initial-value approach (dashed lines). Significant differences between these distributions 
are visible. For helicity-resolved distributions of created electrons (either summed up over the helicities of created positrons for the 
boundary-value- or summed up over the spin degrees of freedom of the initial negative energy states for the initial-value theory), the differences 
are not so pronounced [see, panels (e) and (f) with the same convention concerning the lines]. Still, they differ by four orders of magnitude as compared with the accuracy of our numerical analysis. 
}
\label{figCompareBRUpDown}
\end{figure}

In Fig.~\ref{figCompareBRexK} we compare predictions that follow from solving
the boundary- and initial-value problems. Even though the results look similar, the discrepancies between these two approaches 
exceed our numerical accuracy by four orders of magnitude. Differences are even more pronounced
for the spin-resolved momentum distributions that are shown in Figs.~\ref{figCompareBRUpDown}(a)-(d). This is independently of the fact that 
the partially spin summed up distributions presented in Figs.~\ref{figCompareBRUpDown}(e)-(f) look nearly 
the same. Note, however, that with increasing the electric field strength, frequency or number of pulse cycles, even for the spin summed 
up distributions the discrepancies between the results based on the boundary- and initial-value approaches become significant.

\begin{figure}
\includegraphics[width=7.5cm]{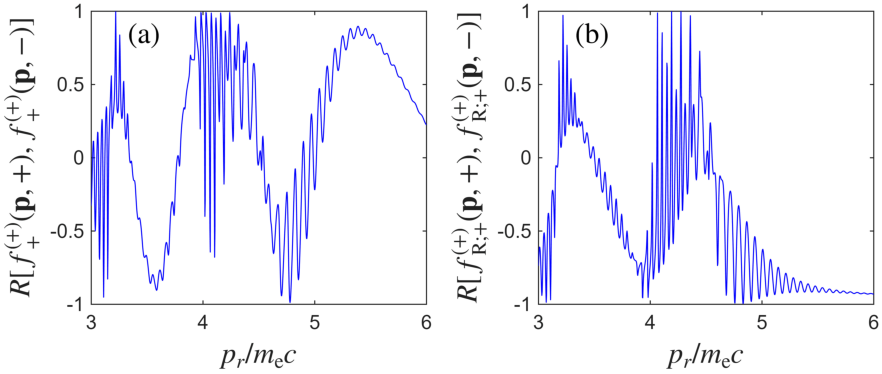}
\caption{Momentum asymmetry distributions [cf., Eq.~\eqref{asy1}] for two selected helicity-resolved momentum distributions obtained by applying either
(a) the boundary-value or (b) the initial-value approaches to the pair creation process. The results are for the electric pulse parameters
presented in Fig.~\ref{fpola}.
}
\label{figCompareRDiff12Kosc11}
\end{figure}

\begin{figure}
\includegraphics[width=7.5cm]{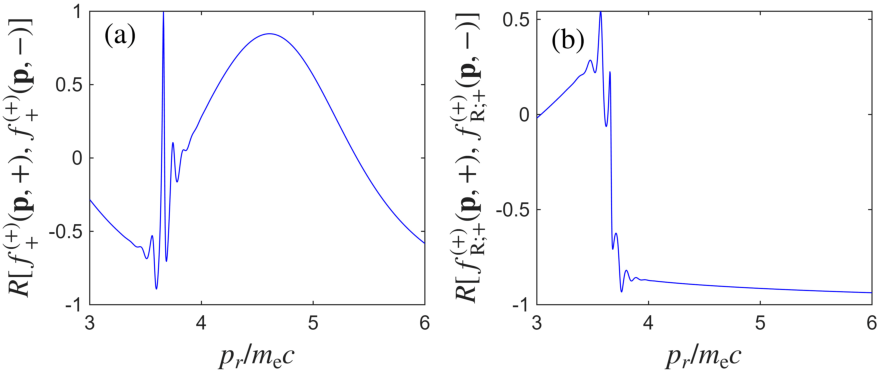}
\caption{Same as in Fig.~\ref{figCompareRDiff12Kosc11} but for the shorter electric pulse, with $N_\mathrm{osc}=5$.
}
\label{figCompareRDiff12Kosc5}
\end{figure}

Both considered approaches permit to determine all four spin-resolved momentum probability distributions for the creation of a single pair 
from the vacuum (for the boundary-value problem) or for the excitation of a single electron from the Dirac sea to the positive energy continuum 
(for the initial-value problem). However, because real experiments (currently hypothetical for $e^-e^+$ pair creation from the vacuum, although already 
being implemented in other areas of physics for analogous processes) generate multiple pairs scaled by unknown factors (like creation volume), 
experimental results are often presented in arbitrary units. Validating the respective theories requires alternative distributions independent 
of these scaling factors. One such possibility consists in using the momentum asymmetry distributions. 
For two different spin-momentum distributions of the same process, $f_1(\bm{p})$ and $f_2(\bm{p})$, they are defined as
\begin{equation}\label{asy1}
R[f_1(\bm{p}),f_2(\bm{p})]=\frac{f_1(\bm{p})-f_2(\bm{p})}{f_1(\bm{p})+f_2(\bm{p})}.
\end{equation}
Some selected asymmetry distributions are presented in Figs.~\ref{figCompareRDiff12Kosc11} and~\ref{figCompareRDiff12Kosc5}, showing significant 
differences between the boundary- and initial-value approaches.

\subsection{Helicity-entangled distributions}\label{sec:Entanglment}

Investigations of quantum entanglement properties and quantum vortex structures in SFQED has been gaining popularity in recent years (see, 
e.g.,~\cite{PhysRevD.95.036006,PhysRevA.99.032340,PhysRevD.109.076004,photonics12040307,9br2-tj4t,k88l-987l,fan2025vortex,jiang2026oam}). 
For this reason, in this section we shall discuss shortly the entanglement of helicity states in the Sauter-Schwinger process, while comparing 
approaches using boundary and initial conditions. However, it should be emphasized that the concept of spin-entanglement has a different meaning 
in these two formalisms. This stems from the Feynman space-time description of QED processes. As discussed above, in the initial-value approach 
there are actually no positrons, but the Dirac sea filled by electrons being in negative energy states (which can also be called the Dirac vacuum).
Only the absence of an electron in such a state is interpreted as a positive energy positron. Therefore, in this case, the pair creation process 
consists in the excitation of an electron from a negative energy state to a positive energy state. The probability amplitude of such an excitation 
is described by Eq.~\eqref{r4}, which contains different time limits. Hence, in the initial-value approach, the spin degrees of freedom refer to 
different asymptotic times. In the boundary-value approach, however, the negative energy solutions of the Dirac equation describe real positive 
energy positrons, and both spin degrees of freedom refer to future times [see, e.g., Eq.~\eqref{ff1}]. Thus, describing the spin states of created pairs.

To begin our analysis of the helicity-entangled momentum distributions let us note that for the homogeneous electric field pulses the amplitudes 
$\mathcal{A}^{(+)}_{S}(\bm{p})$ and $\mathcal{A}^{(+)}_{\mathrm{R};S}(\bm{p})$ vanish. In order to show this, we introduce the notation 
$\mathcal{A}^{(+)}_{S}[\bm{p}|\bm{A}]$ and $\mathcal{A}^{(+)}_{\mathrm{R};S}[\bm{p}|\bm{A}]$ reflecting the fact that probability amplitudes 
also functionally depend on the vector potential $\bm{A}(t)$. Let us restrict our further discussion to the case of 
$\mathcal{A}^{(+)}_{S}[\bm{p}|\bm{A}]$, as the analysis for $\mathcal{A}^{(+)}_{\mathrm{R};S}[\bm{p}|\bm{A}]$ is similar. 
Since $\mathcal{A}^{(+)}_{S}[\bm{p}|\bm{A}]$ does not depend on the choice of the spin quantization axis (as discussed in Sec.~\ref{sec:spin}), 
it is only a function of $\bm{p}^2$ and a functional of $\bm{p}\cdot\bm{A}(t)$ and $\bm{A}^2(t)$, i.e.,
\begin{equation}\label{ent1}
\mathcal{A}^{(+)}_{S}[\bm{p}|\bm{A}]=\mathcal{F}[\bm{p}^2|\bm{p}\cdot\bm{A},\bm{A}^2],
\end{equation}
which means that
\begin{equation}\label{ent2}
\mathcal{A}^{(+)}_{S}[\bm{p}|\bm{A}]=\mathcal{A}^{(+)}_{S}[-\bm{p}|-\bm{A}].
\end{equation}
On the other hand, by applying the space inversion transformation, $\bm{x}\rightarrow -\bm{x}$ and $\bm{A}(t)\rightarrow -\bm{A}(t)$, we find that the Feynman propagator fulfills the equation,
\begin{equation}\label{ent3}
K_{\mathrm{F}}[x^0,\bm{x},y^0,\bm{y}|\bm{A}]=\gamma^0K_{\mathrm{F}}[x^0,-\bm{x},y^0,-\bm{y}|-\bm{A}]\gamma^0,
\end{equation}
where we have introduced the notation indicating that the propagator also depends functionally on $\bm{A}(t)$. Since
\begin{equation}\label{ent4}
u^{(\beta)}_{\bm{p},\lambda}=\beta\gamma^0u^{(\beta)}_{-\bm{p},\lambda},
\end{equation}
it follows from Eq.~\eqref{ff1} that
\begin{equation}\label{ent5}
\mathcal{A}^{(+)}_{S}[\bm{p}|\bm{A}]=-\mathcal{A}^{(+)}_{S}[-\bm{p}|-\bm{A}].
\end{equation}
Hence, together with \eqref{ent2} it indicates that $\mathcal{A}^{(+)}_{S}[\bm{p}|\bm{A}]=0$. In a similar way, we show that 
$\mathcal{A}^{(+)}_{\mathrm{R};S}[\bm{p}|\bm{A}]=0$. We have also checked these results analytically in the first- and second-order Born approximations. 
Our numerical analysis shows that the moduli of both scalar amplitudes are smaller than the assumed numerical precision.

\begin{figure}
\includegraphics[width=7.5cm]{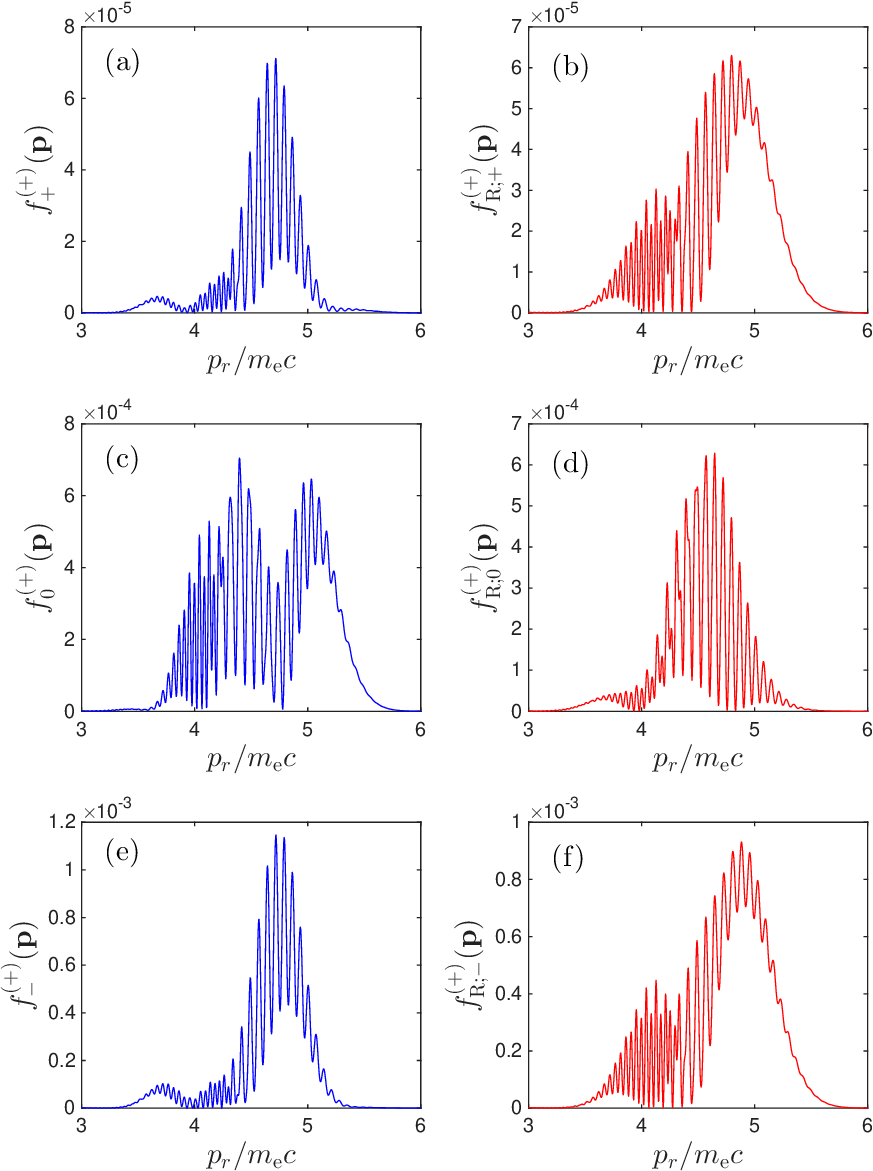}
\caption{Shows helicity-entangled momentum distributions. The left column presents the results of the boundary-value approach [cf., Eq.~\eqref{s5b}] 
and the right column of the initial-value one [cf., Eq.~\eqref{si2b}]. In both cases the spherically symmetric distributions, 
$f^{(+)}_S(\bm{p})$ and $f^{(+)}_{\mathrm{R};S}(\bm{p})$, vanish. As for the spin-resolved distributions significant differences are observed, 
although the total momentum distributions look very similar [cf., Figs.~\ref{figCompareBRUpDown}(e) and~\ref{figCompareBRUpDown}(f)].
}
\label{figCompareBRHelicity}
\end{figure}

In Fig.~\ref{figCompareBRHelicity}, we compare the helicity-entangled momentum distributions by applying the boundary-value approach (left column) 
with the initial-value one (right column). As for the spin-resolved distributions we observe significant differences, even though discrepancies between the helicity summed up distributions are rather small.

In summary, our calculations in Sec.~\ref{sec:num} show that while boundary- and initial-value approaches yield nearly identical spin-averaged results 
for certain pulse parameters, significant discrepancies emerge in spin-resolved and helicity-entangled distributions. At higher intensities and frequencies, these differences persist even in spin-summed 
distributions. Such situation mirrors the tension 
between Sauter and Schwinger theories: while Schwinger's approach aligns with modern QED, Sauter's relies on the Dirac sea, a concept superseded in 
quantum field theory by the Feynman interpretation of positrons, even if both models occasionally produce similar numerical results.

\section{Conclusions}
\label{sec:conclusions}

Different approaches to the Sauter-Schwinger pair creation process are equivalent only if they yield identical results. In this regard, methods 
based on the Dirac equation with initial conditions, such as QKE and DHW formalisms, are equivalent to the modified Feynman space-time 
formalism using retarded or advanced propagators. This equivalence relies on the assumption of a Dirac sea and electron excitation
from negative to positive energy states. Alternatively, methods using the Dirac equation with Feynman or anti-Feynman boundary 
conditions~\cite{bialynicki1975quantumelectro}, the scattering matrix approach, and the original Feynman theory (where negative energy states 
represent antiparticles) are also equivalent. Both classes of approaches yield exact (or "to all orders") solutions to the dynamical Sauter-Schwinger 
problem; however, "exact" does not imply "unique," as there are infinitely many solutions depending on the chosen initial, final, or boundary conditions.
Notably, when investigating the Sauter-Schwinger process in relativistic QED using the original Feynman space-time formalism or scattering matrix 
theory, the boundary-value method must be applied. In contrast, the initial-value methods are applicable for investigating relativistic plasmas 
(see, e.g.,~\cite{brodin2021plasmadynamics,Brodin2022,PhysRevE.107.035204,10.1063/5.0300510}), 
in which the real electrons and positrons already exist, provided however that particles spins do not play a significant role.

Note that the boundary-value method discussed in our work is applicable not only to the relativistic Sauter-Schwinger process, which at the moment is not 
experimentally feasible and theoretical explorations of which are rather concentrated on expected predictions, independently of the 
progress in the generation of very strong electromagnetic background fields in laboratory conditions. In our opinion, 
the boundary-value approach is the only way to investigate correlations of the final electron-hole spin states and their entanglements in condense 
matter physics~\cite{RevModPhys.81.109,PhysRevD.108.116007,PhysRevLett.124.110403,PhysRevD.99.016025}, in which analogues of the Sauter-Schwinger 
process by time-dependent electric field pulses are already experimentally realized. The initial-value approach also provides the method for the 
investigation of spin correlations. However, as follows from this paper, these would be correlations between the spin states of electrons that are defined for different times, 
i.e., in the past for the valence- and in the future for the conduction electrons.

Finally, we conclude our discussion with a quote from the Feynman Lectures on QED 
(Ref.~\cite{feynman1998quantum}, below Eq.~(15-13), with a small modification related to the inhomogeneous Dirac equation and with the proper 
adaptation of symbols): 
\textit{
``$K_{\mathrm{F}}(x_2,x_1)$ has an interpretation consistent with the positron interpretation of negative energy states. Thus when the thing is 
``ordinary'' ($x^0_2 > x^0_1$), an electron is present, and only positive energy states contribute. When the timing is ``reversed'' ($x^0_2<x^0_1$), 
a positron is present, and only negative energy states contribute. On the other hand, $K_{\mathrm{R}}(x_2,x_1)$ does not have so satisfactory an 
interpretation. Although the kernel $K_{\mathrm{R}}(x_2,x_1)$ defined by the same inhomogeneous Dirac equation is also a satisfactory mathematical 
solution, the interpretation of $K_{\mathrm{R}}(x_2,x_1)$ requires the idea of an electron in a negative energy state''.
}
This suggests that the initial-value formalism can be applied in nonrelativistic quantum theories or condense matter physics (in which no antiparticles 
are created, as holes are not antiparticles), but is not applicable to the relativistic QED with the initial vacuum state.  
However, from the purely practical point of view, if the probability of the creation of pairs is small, both the boundary- and initial-value approaches 
provide similar (but not identical) results for some particular momentum distributions, as discussed in this paper.

\section*{Acknowledgements}

We would like to thank P.~Chankowski for discussions concerning Relativistic Quantum Field Theories.

%\input{sf3.bbl}
%\end{document}

%apsrev4-2.bst 2019-01-14 (MD) hand-edited version of apsrev4-1.bst
%Control: key (0)
%Control: author (8) initials jnrlst
%Control: editor formatted (1) identically to author
%Control: production of article title (0) allowed
%Control: page (0) single
%Control: year (1) truncated
%Control: production of eprint (0) enabled
%

\end{document}